\documentclass[aps,prx,longbibliography,reprint,superscriptaddress,dvipsnames]{revtex4-2}
\usepackage[colorlinks=true, allcolors=blue]{hyperref}
\usepackage{amsmath,amssymb,amsfonts,graphicx,tabularx,dcolumn,bm,xcolor,times,physics,xfp}
\newcommand{\nn}{\nonumber}
\newcommand{\bd}{\boldsymbol}

\begin{document}
\title{Majorana-magnon interactions in topological Shiba chains}

\author{Pei-Xin Shen}
\affiliation{International Research Centre MagTop, Institute of Physics, Polish Academy of Sciences, Aleja Lotnikow 32/46, PL-02668 Warsaw, Poland}

\author{Vivien Perrin}
\affiliation{Universit\'e Paris-Saclay, CNRS, Laboratoire de Physique des Solides, 91405, Orsay, France}

\author{Mircea Trif}
\affiliation{International Research Centre MagTop, Institute of Physics, Polish Academy of Sciences, Aleja Lotnikow 32/46, PL-02668 Warsaw, Poland}

\author{Pascal Simon}
\affiliation{Universit\'e Paris-Saclay, CNRS, Laboratoire de Physique des Solides, 91405, Orsay, France}

\date{\today}

\begin{abstract}

A chain of magnetic impurities deposited on the surface of a superconductor can form a topological Shiba band that supports Majorana zero modes and holds a promise for topological quantum computing. Yet, most experiments scrutinizing these zero modes rely on transport  measurements, which only capture local properties. Here we propose to leverage the intrinsic dynamics of the magnetic impurities to access their non-local character. We use linear response theory to determine the dynamics of the uniform magnonic mode in the presence of external $ac$ magnetic fields and coupling the Shiba electrons.  We demonstrate that this mode, which spreads over the entire chain of atoms, becomes imprinted with the parity of the ground state and, moreover, can discriminate between Majorana and trivial zero modes located at the ends of the chain. Our approach offers a non-invasive alternative to the scanning tunneling microscopy techniques used to probe Majorana zero modes. Conversely, the magnons could facilitate the manipulation of Majorana zero modes in topological Shiba chains.

\end{abstract}

\maketitle

\section{Introduction}

There has been a growing interest in detecting and controlling Majorana zero modes (MZMs) in various condensed matter systems, partially driven by their potential for topological quantum computation \cite{Kitaev2001Unpaired,Kitaev2003Faulttolerant,Alicea2012New}.  Many theoretical and experimental efforts have been  made in this direction during the last decades, such as fractional quantum Hall systems \cite{Read2000Paired}, cold atoms \cite{Jiang2011Majorana}, semiconducting nanowires \cite{Lutchyn2010Majorana,Oreg2010Helical}, and topological insulators  with spin-orbit coupling (SOC) proximitized with $s$-wave superconductors (SCs) \cite{Fu2008Superconducting,Fu2009Josephson,Stanescu2010Proximity}. Magnetic atoms (either individually positioned or self-assembled) on top of a conventional superconducting substrate provide one of the most promising platforms for MZMs \cite{Choy2011Majorana,Kjaergaard2012Majorana,Martin2012Majorana,Nadj-Perge2013Proposal,Braunecker2013Interplay,Klinovaja2013Topological,Vazifeh2013SelfOrganized}. Indeed, the resulting in-gap Yu-Shiba-Rusinov (YSR) bound states \cite{Yu1965Bound,Shiba1968Classical,Rusinov1969Superconductivity}  can be used as  building blocks to design spinless $p$-wave SCs with long-range couplings \cite{Pientka2013Topological,Brydon2015Topological,Rontynen2015Topological,Hoffman2016Topological,Menard2015Coherent}.

Although many experiments have reported distinctive transport signatures in the form of a zero-bias conductance peak \cite{Nadj-Perge2014Observation,Ruby2015End,Pawlak2016Probing,Schneider2022Precursors}, which indicates the presence of MZMs, the origin of these zero-bias peaks is still under heavy debate: As trivial zero modes (TZMs) can display the same transport phenomenology for both nanowires \cite{Prada2020Andreev,DasSarma2021Disorderinduced,DasSarma2023Search} and YSR chains \cite{Sau2015Bound,Kim2018Tailoring,Kuster2022NonMajorana}, local probes have difficulty offering unambiguous signatures of MZMs. Therefore, the search for other direct and measurable manifestations of Majorana physics is ongoing.

\begin{figure}
    \centering
    \includegraphics[width=0.98\linewidth]{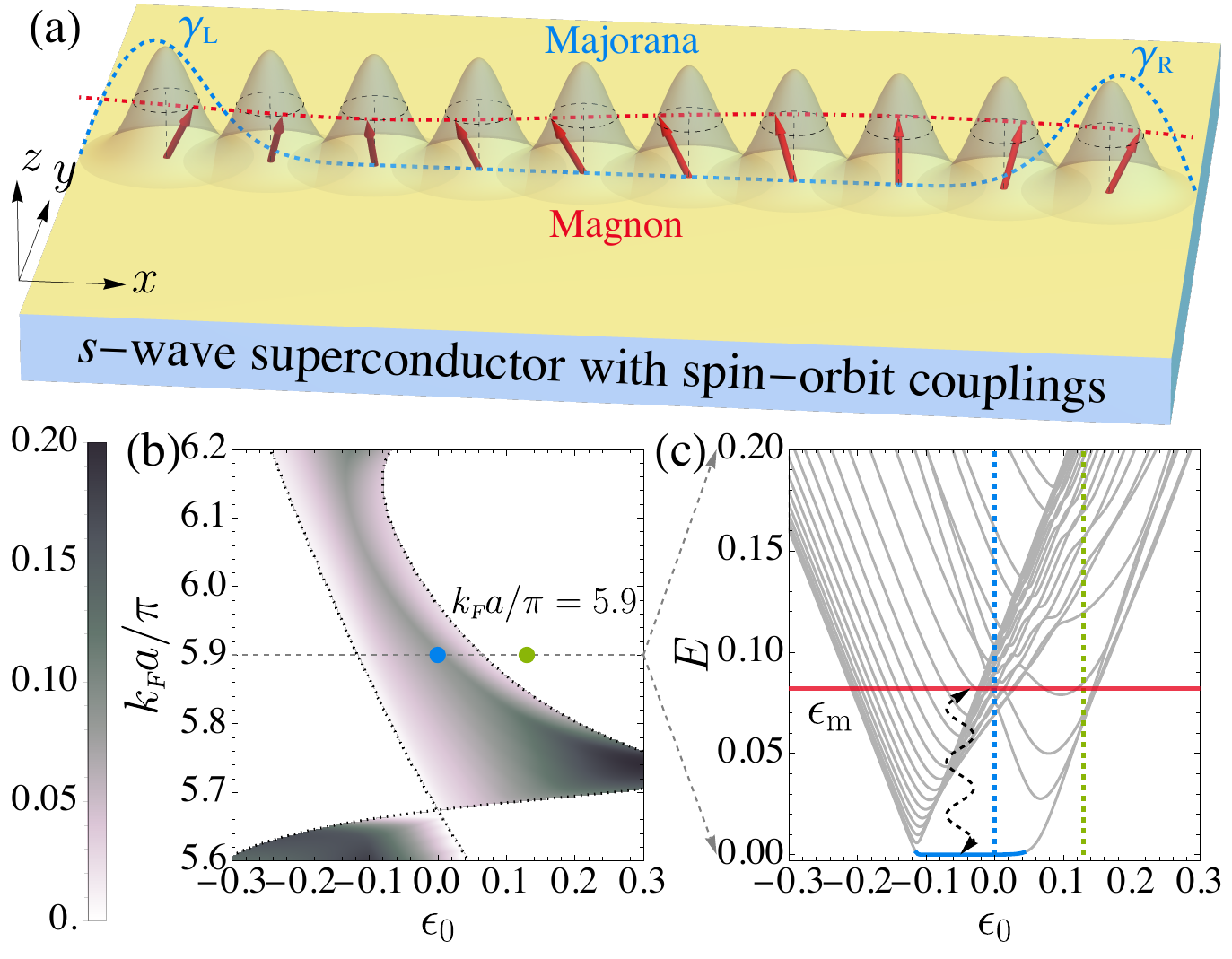}
    \caption{(a) A chain of ferromagnetically coupled adatoms on a two-dimensional $s$-wave superconductor harboring MZMs. The uniform magnonic mode (red dotted line) interacts with the MZMs (blue dashed line), altering its dynamics. (b) Topological phase diagram of the effective one-dimensional model as a function of $k_F a$ and $\epsilon_0$. The dotted lines indicate the boundary between topological (gray shaded) and non-topological (white) phases. The magnitude of the gap can be inferred from the shaded degree. (c) A line cut at $k_F a/\pi = 5.9$, where the blue (green) dashed line corresponds to the blue (green) dot in (b) and lies in a topological (normal) phase. The curved arrow indicates the interaction between the MZM and the uniform magnonic excitation $\epsilon_{\rm m}$. The parameters are $N=30$, $\xi_0 = 10 a$, and $\lambda_R= 0.05 v_F$, where $v_F$ is the Fermi velocity.}
    \label{Fig:Scheme}
\end{figure}

In this paper we propose to harness the  collective spin dynamics of a chain of magnetic impurities deposited on top of a two-dimensional (2D) Rashba $s$-wave SC to both detect MZMs and infer their non-local character. A sketch of the system is shown in Fig.~\ref{Fig:Scheme}(a), depicting the spin-wave excitations (whose quanta are the magnons) interacting with the MZMs. Our approach takes advantage of the inherently strong coupling between the impurities and the superconducting condensate necessary for creating a one-dimensional (1D) topological SC. We demonstrate that the uniform magnonic mode, which can be triggered by external $ac$ magnetic fields, becomes imprinted with the parity of the ground state and, moreover, can discriminate between MZMs and TZMs located at the end of the chain. That is because this mode is extended over the entire length of the chain and therefore susceptible to the non-locality of the MZMs. Its detection can be achieved by ferromagnetic resonance spectroscopic techniques or by coupling it to confined electromagnetic fields in a microwave cavity (see Refs.~\cite{Tabuchi2014Hybridizing,Zhang2014Strongly,Lachance-Quirion2017Resolving,Yuan2022QuantumMagnonics,ZareRameshti2022Cavity,Zheng2023Tutorial} for recent progress and reviews).  

Our findings can be naturally extended to 2D magnetic clusters harboring chiral MZMs \cite{Menard2017Twodimensional,Kezilebieke2020Topological}, superconducting-semiconducting nanowires covered by magnetic insulators \cite{Vaitiekenas2021Zerobias}, or MZM implementations based on carbon nanotubes proximitized by ferromagnets \cite{Desjardins2019Synthetic}, since this proposal  harnesses nonlocal Majorana-magnon coupling and the resulting parity-dependent spin susceptibility.

\section{Model of ferromagnetic Shiba chains}

The Hamiltonian describing a chain of $N$ classical spins coupled to an $s$-wave SC can be written as ~\cite{Pientka2013Topological,Brydon2015Topological,Rontynen2015Topological,Hoffman2016Topological} $H_{\rm tot} = \frac{1}{2} \int \mathrm{d} \bd{r}\hat{\Psi}^{\dagger}(\bd{r})(\mathcal{H}_{\rm el}+\mathcal{H}_{\rm el-m}) \hat{\Psi}(\bd{r})+H_{\rm m}$, with
\begin{align}
    \mathcal{H}_{\rm el} &= \left(\frac{p^2}{2m}-\mu+\lambda_R(p_x\sigma_y-p_y\sigma_x) \right)\tau_z+\Delta \tau_x \,,\nonumber \\
    \label{Eq:Hamiltonian}
    \mathcal{H}_{\rm el-m}&= -J \sum_{j=1}^N \qty(\bd{S}_j \cdot \boldsymbol{\sigma}) \delta(\bd{r} - \bd{r}_j)\,, \\
    H_{\rm m}&=\sum_{\langle i,j\rangle}J_{\rm ex}{\bd S}_i\cdot{\bd S}_j-\sum_{j=1}^N\left(\frac{K_z}{2}(S_{j}^z)^2  - \gamma HS_{j}^z\right)\nonumber\,,
\end{align}
being the pristine spin-orbit coupled SC, its coupling to the impurities with spin size $S=|{\bd S}_j|$, and the bare magnetic Hamiltonian, respectively. Here, $\bd{\tau}=(\tau_x,\tau_y,\tau_z)$ [$\bd{\sigma}=(\sigma_x,\sigma_y,\sigma_z)$] are Pauli matrices acting in the Nambu (spin) space, and $\hat{\Psi}(\bd r)=[\hat{\psi}_{\uparrow}(\bd r), \hat{\psi}_{\downarrow}(\bd r), \hat{\psi}_{\downarrow}^{\dagger}(\bd r),-\hat{\psi}_{\uparrow}^{\dagger}(\bd r)]^\textsc{t}$ is the electronic field operator at position ${\bd r}$. Moreover, $m$ is the electron effective mass, ${\bd p}=-i\hbar\partial_{\bd r}$ is the momentum operator, $\mu$ is the chemical potential, $\Delta$ is the superconducting gap, and $\lambda_R$ is the Rashba SOC strength. Finally, $J$, $J_{\rm ex}$, $K_z$, $\gamma$, and $H$ are the coupling between the spins and the condensate, the (direct) Heisenberg exchange between the spins, the local easy-axis anisotropy, the gyromagnetic ratio, and the applied magnetic field, respectively. Note that the Hamiltonian $\mathcal{H}_{\rm el-m}$ will effectively modify the exchange coupling $J_{\rm ex}$ via the Ruderman-Kittel-Kasuya-Yosida (RKKY) interaction mediated by the SC. This splits into two contributions ($i$) one carried by the bulk SC quasiparticles, analogous to metallic substrates, and ($ii$) one carried by the localized in-gap YSR states that are discussed in the following section. For $k_Fr>\xi_{0}/r$, with $\xi_0$ being the SC coherence length, the former dominates the RKKY exchange coupling \cite{Yao2014Enhanced} (and can determine the magnetic ground state), while the YSR state effects can be accounted for perturbatively, which represents the regime studied in this paper. Motivated by Refs.~\cite{Nadj-Perge2014Observation,Yang2019Coherent,Kuster2022NonMajorana}, we focus on a ferromagnetic alignment of the magnetic impurities either in the plane or perpendicular to the SC surface. 

We assume the magnetic impurities are located at positions $\bd{r}_j=ja\mathbf{e}_x$, with $j=1,\dots,N$ and $a$ the separation between them. They induce $2N$ sub-gap states $\{\phi_j(\bd{r}), \overline{\phi}_j(\bd{r}) \}_{j=1}^N$, where $\phi_j(\bd{r}) = - J_\textsc{e}(\bd{r}-\bd{r}_j)\phi_j(\bd{r}_j)$ is the YSR wavefunction of the $j^{\rm th}$ impurity spinor with $\phi_j(\bd{r}_j) =[1,0,1,0]/\sqrt{\mathcal{N}}$. Here, $\mathcal{N}$ stands for the normalization factor of the YSR wave function shown in Eq.~\eqref{Eq:NormalizationFactor} which plays an important role. Finally,  $J_\textsc{E}(\bd{r})=JS/(2 \pi)^2 \int \mathrm{d} \bd{k} \mathrm{e}^{\mathrm{i} \bd{k} \cdot \bd{r}}(E-\mathcal{H}_\mathrm{el})^{-1}$, while the corresponding hole wavefunction is given by $\overline{\phi}_j(\bd{r}) = \mathcal{C} \phi_j(\bd{r})$, where $\mathcal{C}=\tau_y \sigma_y \mathcal{K}$ is the particle-hole operator and $\mathcal{K}$ is the complex conjugation. Their corresponding energies are $ \epsilon_0 = \pm\Delta (1-\alpha^2)/(1+\alpha^2)$, respectively, where $\alpha = \pi \nu_0 JS$ is the dimensionless impurity strength and $\nu_0$ is the density of states at the Fermi level in the normal state without SOC \cite{Kopnin2001Theory}. For a general state $\psi(\bd{r})$, the Schr{\"o}dinger equation $(\mathcal{H}_\mathrm{el}+\mathcal{H}_\mathrm{el-m}) \psi(\bd{r}) = E \psi(\bd{r})$ can be reduced to a closed set of equations for the spinor at the magnetic impurity positions $(\bd{r}_{ij} = \bd{r}_{i}- \bd{r}_{j})$ \cite{Pientka2013Topological,Brydon2015Topological,Rontynen2015Topological,Hoffman2016Topological}:
\begin{equation}
    [\boldsymbol{S}_i \cdot \bd{\sigma} +J_\textsc{E}(0)] \psi(\bd{r}_i) = -\sum_{j \neq i} J_\textsc{E}(\bd{r}_{ij}) \psi(\bd{r}_j) \,.
    \label{Eq:WaveFunctionLatticeRelationship}
\end{equation}
When the chain is in the deep-dilute limit corresponding to $1/\sqrt{k_F a} \ll 1$ and $\alpha \approx 1$, we can project Eq.~\eqref{Eq:WaveFunctionLatticeRelationship} onto the YSR states and obtain a $2N\times 2N$ effective tight-binding Hamiltonian $\mathcal{H}_\mathrm{eff}$ (see Appendix~\ref{Appendix:EffectiveHamiltonian} for details).  The Hamiltonian $\mathcal{H}_\mathrm{eff}$ belongs to the Altland-Zirnbauer symmetry class D and is characterized by a $\mathbb{Z}_2$ topological invariant \cite{Schnyder2008Classification,Ryu2010Topological,Ludwig2015Topological}. By tuning $k_F a$ and $\epsilon_0$, the system can enter a superconducting topological phase supporting MZMs, whose phase diagram is shown in Fig.~\ref{Fig:Scheme}(b). Specifically, in Fig.~\ref{Fig:Scheme}(c) we show the spectrum of $\mathcal{H}_\mathrm{eff}$ for $N=30$.

\section{Ferromagnetic lattice dynamics}
The magnetic dynamics is dictated by $H_{\rm m}$ with $J_{\rm ex}$ renormalized by the bulk RKKY interaction \footnote{We disregarded the anisotropic Dzyaloshinskii-Moriya interactions stemming from the SOC, which have been found to be negligible in comparison to the Heisenberg RKKY contribution \cite{Kuster2022NonMajorana}. However, we account for its expression in Appendix~\ref{Appendix:LatticeDynamics}.}. We can establish the dispersion of the magnetic fluctuations by employing a Holstein-Primakoff transformation \cite{Holstein1940Field}. In the limit of large $S$, the transformation reads $S_{j}^{+(-)}=\sqrt{2S}a_j(a_j^\dagger)$ and $S_{j}^{z}=S-a_j^\dagger a_j$, with $a_j$ ($a_j^\dagger)$ being the magnonic annihilation (creation) operator satisfying $[a_j,a_{j'}^\dagger]=\delta_{jj'}$. In this paper we are interested in triggering the dynamics of the uniform magnonic mode, $a_0=(1/\sqrt{N})\sum_{j}a_j$ and energy $\epsilon_{\rm m} = K_z S-\gamma H$ because $(i)$ it represents the lowest energy magnon and ($ii$) it exhibits a  constant amplitude along the wire, rendering it highly non-local. This mode isolation applies as long as the energy splitting between the uniform mode and the first excited magnon mode, $\epsilon_{1} - \epsilon_{\rm m} \approx \pi^2 |J_\mathrm{ex}| S / (N+1)^2$, is larger than the magnon decay rate $\kappa_{\rm m}$ (see Appendix~\ref{Appendix:LatticeDynamics}) \cite{Gray2006Toeplitz}. In addition, the magnetic field $H$ is generally close to zero and has few effects on the electrons of the SC. Hence, projecting Eq.~\eqref{Eq:Hamiltonian} onto the uniform mode $a_0$, we find
\begin{equation}
    H^0_{\rm el-m}
    \approx \frac{J}{N}\left[n_0 \Sigma_z - \sqrt{2NS}(a_0^\dagger \Sigma_+ + a_0 \Sigma_-)\right]\,,
    \label{el-m}
\end{equation}
where $\Sigma_\nu\equiv \sum_{j\sigma\sigma'} \hat{\psi}^\dagger_{S\sigma}(\bd{r}_j)\sigma^\nu_{\sigma\sigma'} \hat{\psi}_{S\sigma'}(\bd{r}_j)$ is the total spin operator along the $\nu=x,y,z$ axis stemming from the YSR states, $n_0 = a^\dagger_0 a_0$, $\Sigma_\pm = (\Sigma_x \pm i \Sigma_y)/2$, and $\hat{\psi}_{S\sigma}(\bd{r})\equiv\mathcal{P}_S\hat{\psi}_\sigma(\bd r)\mathcal{P}_S$, where $\mathcal{P}_S$ projects the electronic field operators onto the in-gap YSR states.  

Using the expression for normalization factor ${\mathcal{N}}$ of the YSR wavefunction \cite{Pientka2013Topological} (see Appendix~\ref{Appendix:EffectiveHamiltonian}),
\begin{align}
    \frac{1}{\mathcal{N}}=\frac{\Delta}{JS}\frac{2\alpha^2}{(1+\alpha^2)^2}\,,
    \label{Eq:NormalizationFactor}
\end{align}
effectively entails substituting $J\rightarrow\Delta/S$ and $\Sigma_\nu\rightarrow 2\mathcal{N}\alpha^2(1+\alpha^2)^{-2}\Sigma_\nu \equiv \widetilde{\Sigma}_\nu$ in Eq.~\eqref{el-m}. This in turn implies that the electronic response is governed by the low-energy scale $\Delta$ ($\alpha\approx 1$) instead of the coupling strength $J$ as could have been anticipated from the prefactor of Eq.~\eqref{el-m}. To determine the effect of the YSR states on the uniform magnon mode, we evaluate the response to a uniform $ac$ magnetic field $h(t)=h_0\mathrm{e}^{-\mathrm{i}\omega t}$ applied perpendicular to the magnetization, with $h_0$ and $\omega\sim\epsilon_{\rm m}$ being its amplitude and frequency, respectively. Following  Refs.~\cite{Dmytruk2015Cavity,Trif2019Braiding,Aseev2019Degeneracy,Shen2021Theory}, we can exploit the input-output theory to quantify the magnonic response in leading order in the electron-magnon interaction. The amplitude of this uniform mode in the frequency space becomes
\begin{align}
    a_0(\omega)=
    \frac{ih_0}{\omega-[\epsilon_{\rm m}+ \frac{\Delta}{NS} \langle\widetilde{\Sigma}_z\rangle+ \frac{2\Delta^2}{NS} \Pi_{+-}(\epsilon_{\rm m})] + i\kappa_{\rm m}}\,,
\end{align}
where 
\begin{align}
\label{BareSusceptibility}
    \Pi_{+-}(\omega)=-i \int_{-\infty}^{\infty}\mathrm{d} t \mathrm{e}^{ i\omega t} \theta(t)\langle[\widetilde{\Sigma}_+(t),\widetilde{\Sigma}_-(0)]\rangle \,,
\end{align}
is the transverse susceptibility associated with the operator $\widetilde{\Sigma}_{\pm}$.  Here, $\expval{\dots}$ represents the expectation value over the electronic density matrix $\rho_{\rm el}$ in the absence of magnons. Therefore, the magnon resonance frequency and its decay are respectively shifted by 
\begin{align} \delta\epsilon_{\rm m}&=\frac{\Delta}{NS}\left(\langle\widetilde{\Sigma}_z\rangle+2\Delta \Re\Pi_{+-}(\epsilon_{\rm m})\right)\nonumber\,,\\
    \delta\kappa_{\rm m}
    &=-\frac{2\Delta^2}{NS}\Im\Pi_{+-}(\epsilon_{\rm m})\,,
\end{align}
which represents one of our main results. The magnitude of these changes is determined by the SC gap $\Delta$ reduced by the total number of spins in the chain $NS$. In order to evaluate them explicitly, we need to specify the density matrix $\rho_{\rm el}$. At zero temperature (or temperatures much smaller than the topological gap) and in the topological regime, the two lowest many-body states  are spanned by $|0\rangle=\ket{\rm vac}$ and $|1\rangle=d_0^\dagger \ket{\rm vac}$, where the full fermion operator $d_{0}$ satisfies $d_0|{\rm vac}\rangle=0$ and can be decomposed into two Majorana operators $\gamma_L,\gamma_R$ (localized on the left and right edge of the chain) as  $d_0\equiv\gamma_L+i\gamma_R$ \cite{Alicea2011NonAbelian} . These two many-body states, which become degenerate in the limit of large $N$, correspond to the even ($\mathcal{P}=0$) and odd ($\mathcal{P}=1$) parity, respectively, and could serve to encode a topologically protected qubit.  In the following, we assume a density matrix of definite parity, $\rho_{\rm el}^{\mathcal{P}}=|\mathcal{P}\rangle\langle\mathcal{P}|$, and that the measurement is performed on time scales much shorter than the time it takes to reach a mixed state via, for example, quasiparticle poisoning. 

\begin{figure}[t]
    \centering
    \includegraphics[width=\linewidth]{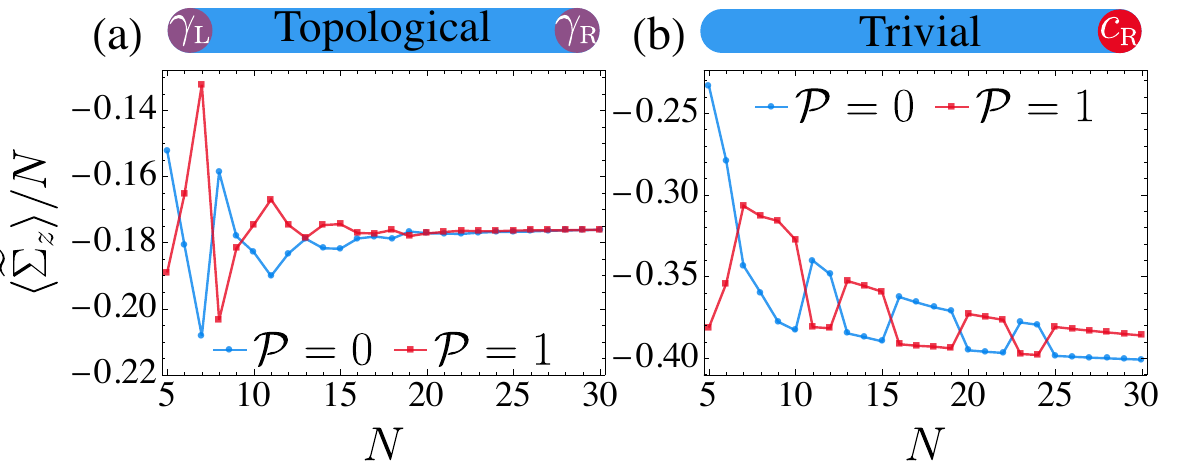}
    \caption{MZM vs. TZM spin expectation values. In (a) and (b) are average spin expectation values along the $z$ axis for the topological phase and the trivial phase, respectively, as a function of the total impurity spins $N$. As the number of sites increases, $\langle\widetilde{\Sigma}_z \rangle/N$ is different for different parities in the trivial regime ($\epsilon_0 = +0.13$) while it converges to the same value in the topological regime ($\epsilon_0 = 0$). The blue (red) lines refer to the $\mathcal{P}=0(1)$ parity.}
    \label{Fig:TotalSpin}
\end{figure}

In Fig.~\ref{Fig:TotalSpin}(a) we show the expectation value of the average spin $\langle\widetilde{\Sigma}_z\rangle/N$ for the two parities, as a function of the number of impurities in the chain. While each parity exhibits different values of $\langle\widetilde{\Sigma}_z\rangle/N$ when $N$ is small, they  become exponentially indiscernible for $Na\gg\xi_0$. Hence, the two parities cannot be discriminated by $\langle\widetilde{\Sigma}_z\rangle/N$ alone, as expected for well-separated MZMs \cite{Ben-Shach2015Detecting}. This exponential sensitivity of the magnon frequency shift on the MZMs separation could be exploited to determine the Majorana fusion rules in future experiments, in analogy with the proposals in semiconducting nanowires that utilize charge sensing instead \cite{Aasen2016Milestones}. %

It is instructive to contrast this case with that of a TZM located at one end of the chain. We have fine tuned the exchange coupling strength $J$ of the last impurity in the trivial phase, and we depict in Fig.~\ref{Fig:TotalSpin}(b)  one representative situation which would exhibit a zero-bias-conductance peak in tunneling experiments. We see that $\langle\widetilde{\Sigma}_z\rangle$ is different for the two parities even for a large number of impurities, $Na\gg\xi_0$. 

\section{Spin susceptibility and Robustness}

Next we focus on the evaluation of the spin susceptibility. In the presence of MZMs, this is determined by transitions between the bulk (or extended) levels, as well as between bulk and the MZMs. Since the former is independent of the parity $\mathcal{P}$, we focus only on the latter, which dominates the susceptibility in the regime $\Delta_{\rm eff}<\epsilon_{\rm m}<2\Delta_{\rm eff}$ \cite{Dmytruk2015Cavity}. However, for the sake of completeness, in Appendix~\ref{Appendix:Susceptibility} we put forward the susceptibility for arbitrary $\epsilon_{\rm m}$.  The susceptibility involving the MZMs reads
\begin{align} 
    \label{Pi_nM=1}
    \Pi_{+-}(\omega,\mathcal{P})=-
    \sum_{E_n>0}\frac{(-1)^\mathcal{P}\mathcal{O}^{\mathcal{P}+}_{0n}\mathcal{O}^{\mathcal{P}-}_{n0}}{\omega-E_n-(-1)^\mathcal{P}E_0+i\eta}\,,
\end{align}
where \footnote{We remark that one should take into account all components of $J_\textsc{E}(\bd{r})$ given in Ref.~\cite{Brydon2015Topological} to compute the susceptibility. The approximated form of $J_\textsc{E}(\bd{r})$ given in Ref.~\cite{Rontynen2015Topological} will result in a zero susceptibility.}
\begin{align}
    \mathcal{O}^{\mathcal{P}\pm}_{nm}
    &= \sum_j[ \Phi_n^\dagger(\bd{r}_j)\delta_{\mathcal{P}1}+\overline{\Phi}_n^\dagger(\bd{r}_j)\delta_{\mathcal{P}0}]\sigma_\pm\Phi_m(\bd{r}_j)\,, 
\end{align}
is the corresponding parity-dependent matrix elements \footnote{Because $\sigma_\pm$ are not Hermitian, the  matrix element $\mathcal{O}^{0-}_{nm}$ should be evaluated as $ (\mathcal{O}^{0+}_{nm})^*$. The details can be found in Appendix~\ref{Appendix:Susceptibility}} and $\Phi_n(\bd{r})$ is the wavefunction pertaining to the Bogoliubon with energy $E_n$ (see Appendix~\ref{Appendix:Transformation}). Fig.~\ref{Fig:Susceptibility} (a) and (b) shows the parity dependent magnon absorption as a function of the driving frequency $\omega$ for MZMs and TZMs, respectively (same as in Fig.~\ref{Fig:TotalSpin}). While both exhibit a peak structure because of the resonances at $\omega=E_n-E_0$, their distinction is encoded in their amplitudes. To quantify it, we define the visibility of the spin susceptibility associated with the two parities as follows:
\begin{equation}
    \label{Eq:Visibility}
    \mathcal{V}(\omega) \equiv \frac{\Im\Pi_{+-}(\omega,0)-\Im\Pi_{+-}(\omega,1)}{\Im\Pi_{+-}(\omega,0)+\Im\Pi_{+-}(\omega,1)} \,,
\end{equation}
which is shown in Fig.~\ref{Fig:Susceptibility} (c) and (d) for the MZMs and the TZMs, respectively. We see that $\mathcal{V}(\omega)$ oscillates between $-1$ and $1$ in the topological regime, while it does not in the trivial regime. This significant difference can be traced back to the symmetry of the pristine 1D system: the effective Hamiltonian $H_{\rm eff}$ is invariant under the symmetry operation $\mathcal{S}=\tau_z\otimes\mathcal{I}$, where $\mathcal{I}$ is the inversion operator that maps site $j$ into $N+1-j$. Hence, the $n^{\rm th}$ single-particle eigenstate is either symmetric or anti-symmetric under $\mathcal{S}$, corresponding to the eigenvalues  $S_n=1$ and $-1$, respectively.  This further reflects onto the transitions matrix elements which satisfy (see Appendix~\ref{Appendix:QuantizedVisibility})
\begin{figure}[t]
    \centering
    \includegraphics[width=\linewidth]{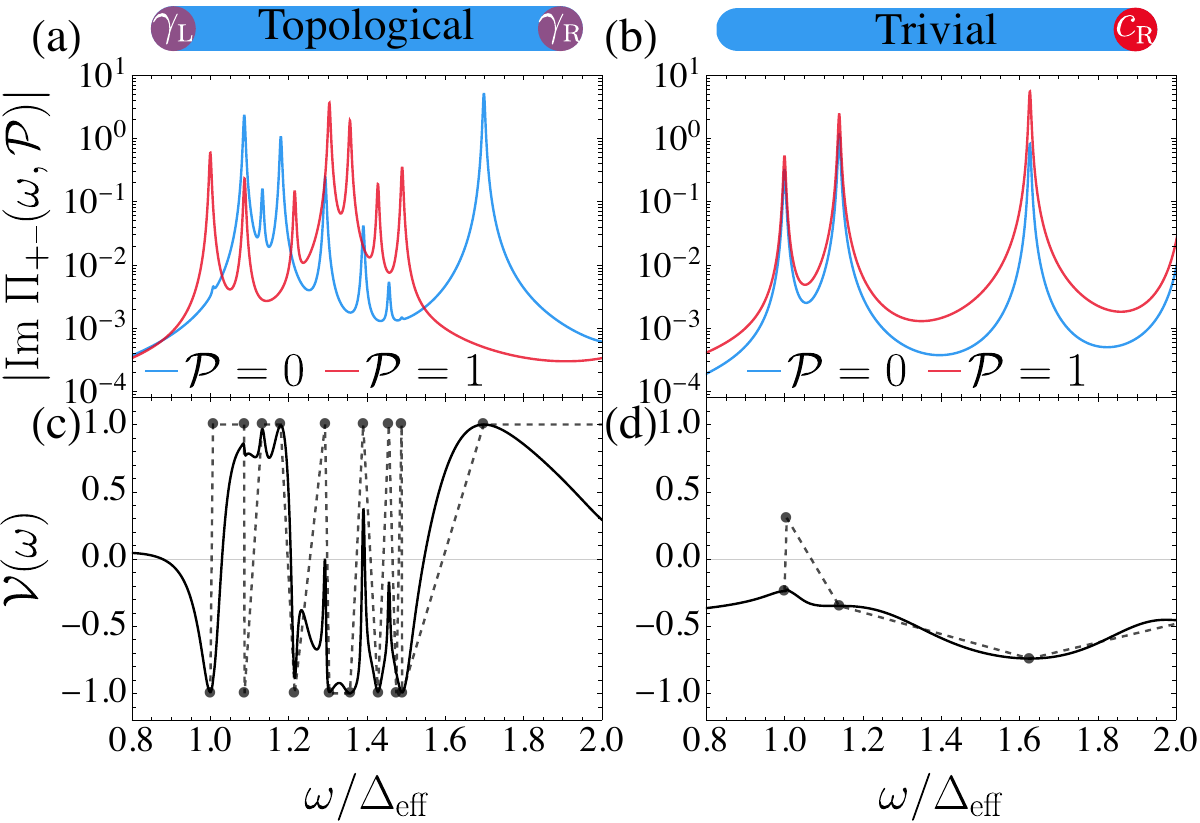}
    \caption{Frequency dependence of the susceptibility for $N=30$. The curves in panels (a) and (b) are the imaginary part of the parity-dependent spin susceptibility for topological and trivial zero energy end modes, respectively. The solid lines in panels (c) and (d) represent the visibility of spin susceptibility defined in Eq.~\eqref{Eq:Visibility}. The full circles connected by dashed lines pertain to the resonances $\omega=E_n-E_0$ in the limit $\eta\rightarrow0$. The visibility oscillates between $-1$ and $1$ in the topological regime, while it takes arbitrary values for accidental zero modes located at the wire ends in the trivial regime. All energies are expressed in terms of the topological gap $\Delta_\mathrm{eff}$.}
    \label{Fig:Susceptibility}
\end{figure}
\begin{align}
    \mathcal{O}^{\mathcal{P}\pm}_{0n}=(-1)^{\mathcal{P}}S_{0}S_n\mathcal{O}^{\mathcal{P}\pm}_{0n}\,.
    \label{symmetry}
\end{align}
Therefore, $\mathcal{O}^{\mathcal{P}\pm}_{0n}\neq0$ only when $(-1)^\mathcal{P}S_0S_n=1$, which means one of the parities always gives a vanishing contribution for any transition. The amplitude of the visibility at the resonances $\omega_n\equiv E_n-E_0$ in the limit $\eta\rightarrow0$ becomes $\mathcal{V}(\omega_n)=S_0S_n\equiv\pm1$, which is depicted by black dots in Fig.~\ref{Fig:Susceptibility}(c). On the other hand, an accidental zero-energy mode located at one edge in the trivial regime severely breaks the inversion symmetry, rendering the visibility arbitrary [Fig.~\ref{Fig:Susceptibility}(d)]. This behavior is also intimately related to the non-locality of the MZMs (as opposed to the locality of the TZMs): they are sensitive to the entire wire because the bulk states excited by the uniform magnonic mode need to travel between the two ends in order to discriminate between the two parities (they are mostly sensitive to the region around their position).

\begin{figure}[t]
    \centering
    \includegraphics[width=\linewidth]{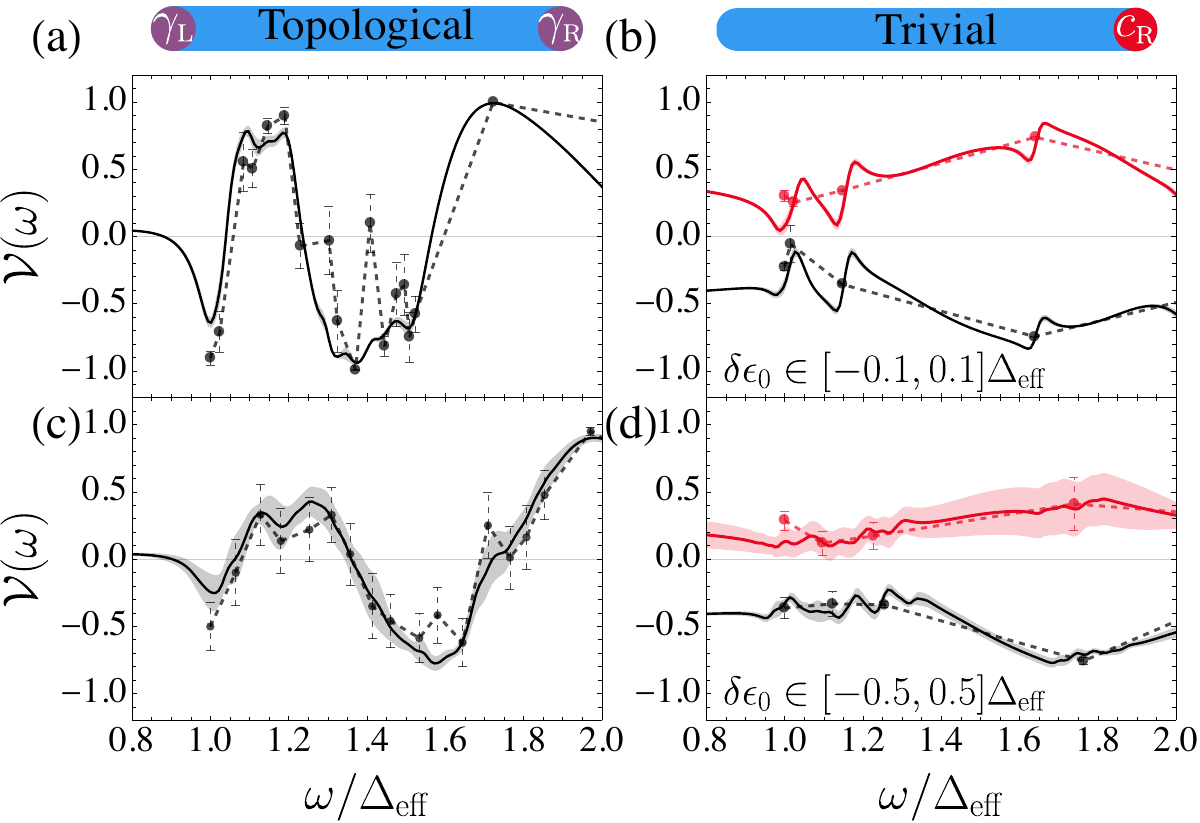}
    \caption{Frequency dependence of the visibility $\mathcal{V}(\omega)$ in the presence of random onsite disorder $\delta\epsilon_0$. The subplots in the left and right columns are in the topological and trivial regimes, respectively. In panel (a) and (b) we consider uncorrelated uniformly distributed random energies $\delta\epsilon_0$ between $[-0.1, 0.1]\Delta_{\rm eff}$, while in (c) and (d) between $[-0.5, 0.5]\Delta_{\rm eff}$. The red curves pertain to disorder realizations that push the TZM slightly below the zero energy,  and thus the sign of the visibility is inverted (because of the parity flip). The solid lines and full circles are the mean of ten realizations, whose standard errors are indicated in the shaded area and error bars, respectively. All other parameters are the same as in Fig.~\ref{Fig:Susceptibility}. }
    \label{Fig:Disorders}
\end{figure}

To test how deviations from the pristine inversion symmetry alter the visibility $\mathcal{V}(\omega)$, we have added random disorder in the individual Shiba energy $\epsilon_0$ along the chain. In Fig.~\ref{Fig:Disorders} 
 we show the visibility for the MZMs (left column) and the TZMs (right column), respectively, for several disorder realizations and strengths. We see that the oscillation of the visibility  remains intact for the MZMs, albeit with a reduced amplitude. The TZMs' visibility, on the other hand, is practically unaffected by disorder because they are local and therefore insensitive to the interference pattern of the bulk modes. 

The linewidth of the bulk Shiba levels, $\eta$, can also affect the visibility.  Indeed, $\delta\Pi_{+-}(\omega)=|\Pi_{+-}(\omega,0)-\Pi_{+-}(\omega,1)|\propto1/N$ in the ballistic regime,  which  persists as long as the average energy level spacing of the bulk levels ($\delta\epsilon$) satisfies $\delta\epsilon\approx v_F/Na>\eta$. This in turn can be associated with a chain length $N^*=v_F/a\eta$,  beyond which $\delta\Pi_{+-}(\omega)\sim e^{-N/N^*}$. This can be interpreted as follows: a bulk state injected at the left side of the chain has its amplitude reduced once reaching the other chain end because of its finite lifetime, therefore diminishing its common overlap with the two MZMs.

\section{Conclusions and outlook}
In this paper we have studied the interaction between the MZMs and magnons in ferromagnetically aligned magnetic impurities coupled to a SOC $s$-wave superconductor. We unravelled the non-local MZMs imprints onto the uniform magnonic mode, and demonstrated  their intimate connection with the spatial symmetry of the chain. Finally, we discriminated the effect of MZMs and TZMs from the magnonic response and showed the robustness of the response against moderate onsite disorder. 

There are several possible future directions. By extending to 2D impurity implementations, it should be possible to interface magnons with chiral Majorana modes  \cite{Rontynen2015Topological, Heimes2015Interplay}. Additionally, it could be beneficial to use the magnonic mode actively for processing quantum information with MZMs \cite{Mishra2021YuShibaRusinov,Contamin2021Hybrid,Kornich2021Braiding}. Further down the road, it would be interesting to extend current machine learning techniques to detect topological structures based on the data of spin susceptibility \cite{Tibaldi2023Unsupervised,Yu2023Unsupervised}.

\acknowledgments

We acknowledge helpful discussions with Archana Mishra, Thore Posske, Li-Wei Yu, and Fang-Jun Cheng. MT thanks Jose Lado for his warm hospitality at Aalto University where this work was finalized. PS would like to acknowledge support from the French Agence Nationale de la Recherche (ANR), under Grant No.~ANR-22-CE30-0037-02, as well as Maxime Garnier and Andrej Mesaros for collaborations on related questions at an early stage of this work. This work is supported by the Foundation for Polish Science through the international research agendas program co-financed by the European Union within the smart growth operational program, and by the National Science Centre (Poland) OPUS Grant No.~2021/41/B/ST3/04475. The source code for some analytical and numerical implementations can be found at Ref.~\cite{Note10}.

P.-X.S. and V.P. contributed equally to this work.

\urlstyle{same}
\footnotetext[10]{The source code is available at \url{https://github.com/peixinshen/MajoranaMagnonYSR}}

\appendix
\begin{widetext}

\section{The effective Hamiltonian of the Yu-Shiba-Rusinov chain}
\label{Appendix:EffectiveHamiltonian}

The technical details of the effective tight-binding model have been extensively studied in Refs.~\cite{Pientka2013Topological,Brydon2015Topological,Rontynen2015Topological,Hoffman2016Topological}; here we only sketch the main procedures. Without magnetic couplings, the Hamiltonian of the system contains two parts ${H}_{\rm el} + {H}_{\rm el-m} = \frac{1}{2} \int \mathrm{d} \bd{r}\hat{\Psi}^{\dagger}(\bd{r}) (\mathcal{H}_{\rm el} + \mathcal{H}_{\rm el-m}) \hat{\Psi}(\bd{r})$, with the Bogoliubov-de Gennes Hamiltonian:
\begin{equation}
    \mathcal{H}_{\rm el} = \tau_z \otimes \left(\varepsilon_{\bd{p}} \sigma_0+\bd{l}_{\bd{p}} \cdot \boldsymbol{\sigma}\right)+\Delta \tau_x \otimes \sigma_0\,, \quad
    \mathcal{H}_{\rm el-m}= -J \sum_{j=1}^N \tau_0 \otimes \qty(\bd{S}_j \cdot \boldsymbol{\sigma}) \delta(\bd{r} - \bd{r}_j)\,,
\end{equation}
where $\varepsilon_{\bd{p}} = p^2/2m - \mu$ is the non-interacting dispersion, $\bd{l}_{\bd{p}} = \lambda_R (p_y \mathbf{e}_x - p_x \mathbf{e}_y) = p \lambda_R ( \sin \theta \mathbf{e}_x - \cos \theta \mathbf{e}_y)$ quantifies the Rashba spin-orbit coupling, and the impurity spin $\bd{S}_j=S\left(\sin \zeta_j \cos \vartheta_j, \sin \zeta_j \sin \vartheta_j, \cos \zeta_j\right)$ is parametrized with angle $\zeta_j, \vartheta_j$ around the $y,z$ axis at site $j$. The electronic Green's function is $G(E)= 1/ (E -  \mathcal{H}_{\rm el}) = \left[G_{+}(E)+G_{-}(E)\right]/2$ with
\begin{equation}
    \label{Eq:GreenFunction}
    G_{\pm}(E)=\frac{\left(E \tau_0+\varepsilon_{\pm} \tau_z+\Delta \tau_x\right) \otimes \left(\sigma_0 \pm \sin \theta \sigma_x \mp \cos \theta \sigma_y\right)}{E^2-\varepsilon_{\pm}^2-\Delta^2} \,,
\end{equation}
where  $\varepsilon_{\pm} = \varepsilon_{\bd{p}} \pm |\bd{l}_{\bd{p}}| = \varepsilon_{\bd{p}} \pm p \lambda_R$ are the dispersions of two helicity bands. In the absence of superconductivity, the Rashba spin-orbit coupling lifts the spin degeneracy of the 2D bulk, giving rise to two distinct Fermi momenta $k_F^\pm = k_F\left(\sqrt{1+\lambda^2} \mp \lambda\right)$ and the densities of states at the Fermi level $\nu_\pm = \nu_0 [1 \mp \lambda / \sqrt{1 + \lambda^2}]$, where $\lambda= \lambda_R /v_F$ is a dimensionless spin-orbit strength, $v_F = (\hbar k_F/m) \sqrt{1 + \lambda^2}$ is the Fermi velocity, and $\nu_0 = {m}/(2\pi \hbar^2)$ is the density of states without spin-orbit coupling \cite{Brydon2015Topological}. We begin with the equation $\ket{\psi} = G(E) \mathcal{H}_\mathrm{el-m} \ket{\psi}$ and project the wave function on the position $\bd{r}$:
\begin{align}
    \kern-0.5em
    \psi(\bd{r}) 
    \equiv \bra{\bd{r}}\ket{\psi} 
    &= - \sum_{j=1}^N \left[  JS \int \frac{\mathrm{d} \bd{k}}{(2\pi)^2} G(E) \mathrm{e}^{\mathrm{i} \bd{k} \cdot (\bd{r}-\bd{r}_j) } \right] \tau_0\otimes(\mathbf{e}^\mathbf{s}_j \cdot \bd{\sigma}) \expval{\bd{r}_j|\psi} \equiv - \sum_{j=1}^N J_\textsc{e}(\bd{r} - \bd{r}_j) \tau_0\otimes(\mathbf{e}^\mathbf{s}_j \cdot \bd{\sigma}) \psi(\bd{r}_j) \,.
    \label{Eq:WaveFunctionOverSpace}
\end{align}
For a 2D lattice, the integral $J_\textsc{e}(\bd{r})$ has terms proportional to $\sigma_x$ \cite{Rontynen2015Topological}. Yet for the 1D chain with impurities deposited along the $x$ axis ($\bd{r} = x \mathbf{e}_x$), those terms are canceled out, and thus $J_\textsc{e}(\bd{r})$ is left with the following six terms \cite{Brydon2015Topological}: 
\begin{align}
J_\textsc{e}(\bd{r}) 
    &= \frac{J S}{2}\Big\{\left[I_3^{-}(x)+I_3^{+}(x)\right] \left( E {\tau}_0 \otimes {\sigma}_0+\Delta {\tau}_x \otimes {\sigma}_0\right) + \left[I_1^{-}(x)+I_1^{+}(x)\right] {\tau}_z \otimes {\sigma}_0 \nn\\
    &\qquad \;\; +\left[I_4^{-}(x)-I_4^{+}(x)\right] \left(E {\tau}_0 \otimes {\sigma}_y+\Delta {\tau}_x \otimes {\sigma}_y\right)+\left[I_2^{-}(x)-I_2^{+}(x)\right] {\tau}_z \otimes {\sigma}_y \Big\} \,.
    \label{Eq:Jmatrix}
\end{align}
Note that $ I^\pm_{3}(0) = - \pi \nu_\pm/\sqrt{\Delta ^2-E ^2}$, while $I^\pm_{1,2,4}(0) = 0$, whose explicit expressions as a function of a general position are shown in Eq.~\eqref{Eq:IntegralAsymptotic}. We set $\bd{r} = \bd{r}_i$ and move the $j=i$ term in Eq.~\eqref{Eq:WaveFunctionOverSpace} to the left-hand side $(\bd{r}_{ij} = \bd{r}_{i}- \bd{r}_{j})$:
\begin{equation}
    \label{Eq:WaveFunctionSiteRelationship}
    \left[1 + J_\textsc{e}(0) \tau_0\otimes(\mathbf{e}^\mathbf{s}_i \cdot \bd{\sigma}) \right] \psi(\bd{r}_i)
    = - \sum_{j\neq i} J_\textsc{e}(\bd{r}_{ij}) \tau_0\otimes(\mathbf{e}^\mathbf{s}_j \cdot \bd{\sigma}) \psi(\bd{r}_j) \,.
\end{equation}
The right-hand side is zero when there is only one impurity. In this special case $\mathcal{H}^{N=1}_\mathrm{el-m} = -J \bd{S}_0 \cdot \boldsymbol{\sigma} \delta(\bd{r} - \bd{r}_0)$, we evaluate $J_\textsc{e}(0) = -\alpha/\sqrt{\Delta ^2-E ^2} (E \tau_0 + \Delta \tau_x)$ in the left-hand side and obtain two sub-gap solutions for $E = \pm \epsilon_0 \equiv \pm \Delta (1-\alpha^2)/(1+\alpha^2)$, whose spinors living in the space spanned by the Nambu and spin space are shown as,
\begin{equation}
    \label{Eq:ShibaBasis}
    \phi_0 \equiv \psi(\bd{r}_0) = \frac{\ket{+} \ket{\uparrow}}{\sqrt{\mathcal{N}}}=
    \frac{1}{\sqrt{\mathcal{N}}}
    \begin{bmatrix}
        1 \\ 1 
    \end{bmatrix} \otimes
    \begin{bmatrix}
        \cos (\zeta / 2) \\ \mathrm{e}^{+\mathrm{i} \vartheta} \sin (\zeta / 2)
    \end{bmatrix}
    \, ; \quad 
    \overline{\phi}_0 \equiv \overline{\psi}(\bd{r}_0) = \frac{\ket{-}\ket{\downarrow}}{\sqrt{\mathcal{N}}}=
    \frac{1}{\sqrt{\mathcal{N}}}
    \begin{bmatrix}
        1 \\ -1 
    \end{bmatrix} \otimes
    \begin{bmatrix}
        -\mathrm{e}^{-\mathrm{i} \vartheta} \sin (\zeta / 2) \\ \cos (\zeta / 2)
    \end{bmatrix} \;.
\end{equation}
where $\overline{\phi}_0 = \mathcal{C}\phi_0$ is the hole wavefunction, and $\mathcal{C}=\tau_y \sigma_y \mathcal{K}$ is the particle-hole operator under the Nambu space. The normalization factor $\mathcal{N}$ of these YSR states is [see Eq.~\eqref{Eq:Normalization} for details]
\begin{equation}
    \mathcal{N}
    = \frac{(1+\alpha^2)^2}{2 \pi \nu_0 \alpha \Delta} = \frac{JS}{\Delta} \frac{(1+\alpha^2)^2}{2 \alpha^2} \,.
\end{equation}
Such a normalization factor $\mathcal{N}$ is an important quantity as it weighs the contribution of YSR states to the final observables.

For a ferromagnetic chain with $N$ adatoms, there are $2N$ YSR states forming a band structure within the  gap.
We expand the right-hand side in Eq.~\eqref{Eq:WaveFunctionSiteRelationship} around $E=0$, $\alpha \approx 1$, where $\tau_0\sigma_0$ and $\tau_0\sigma_y$ terms in $J_\textsc{e}(\bd{r})$ are zero, and project the wave function into a set of local Shiba bases $\psi(\bd{r}_j) = p(j)\phi_j(\bd{r}_j)  + q(j) \overline{\phi}_j(\bd{r}_j)$. Comparing the coefficients of the basis, we obtain \cite{Brydon2015Topological}
\begin{align}
\mathcal{H}_\mathrm{eff}
\varphi &= 
\left(
\begin{bmatrix}
    h_\mathrm{eff} &\Delta_\mathrm{eff}\\
    \Delta_\mathrm{eff}^\dagger&-h^*_\mathrm{eff}
\end{bmatrix} + \tau_0 \otimes b_\mathrm{eff}
\right)
\begin{bmatrix}
    p \\
    q
\end{bmatrix} = E
\begin{bmatrix}
    p \\
    q
\end{bmatrix} \,,\quad \text{where} \quad
\begin{cases}
    p = [p(1), \dots, p(j), \dots, p(N)]^\textsc{t}\\
    q = [q(1), \dots, q(j), \dots, q(N)]^\textsc{t}
\end{cases} \,, \nn \\
h_\mathrm{eff} &= + \epsilon_0 \delta_{ij} + \frac{J S \Delta^2}{2} \lim_{E\rightarrow 0} \left[I_3^{-}\left(x_{i j}\right) + I_3^{+}\left(x_{i j}\right)\right]\,, \quad b_\mathrm{eff} = + \frac{J S \Delta^2}{2} \sin(\zeta) \sin (\vartheta) \lim_{E\rightarrow 0} \left[I_4^{-}\left(x_{i j}\right) - I_4^{+}\left(x_{i j}\right)\right] \,, \nn\\
\Delta_\mathrm{eff} &= - \mathrm{i} \frac{J S \Delta}{2} [\cos^2(\zeta/2) + \sin^2(\zeta/2) \mathrm{e}^{-2 \mathrm{i} \vartheta}] \lim_{E\rightarrow 0} \left[I_2^{-}\left(x_{i j}\right)-I_2^{+}\left(x_{i j}\right)\right] \,.
\label{Eq:HeffTwoAngles}
\end{align}
Specifically for the impurities polarized in the $x$-$z$ plane, we have $\vartheta=0$ and $b_\mathrm{eff}=0$, and $\mathcal{H}_\mathrm{eff}$ is reduced to 
\begin{align}
    h_\mathrm{eff} = \epsilon_0 \delta_{ij} + \frac{J S \Delta^2}{2} \lim_{E\rightarrow 0} \left[I_3^{+}\left(x_{i j}\right) + I_3^{-}\left(x_{i j}\right)\right] \,, \quad
    \Delta_\mathrm{eff} = \mathrm{i} \frac{J S \Delta}{2} \lim_{E\rightarrow 0} \left[I_2^{+}\left(x_{i j}\right) - I_2^{-}\left(x_{i j}\right)\right] \,,
    \label{Eq:HeffPlane}
\end{align}
which is independent of the angle $\zeta$ that winds around the $y$ axis. The YSR states in Eq.~\eqref{Eq:ShibaBasis} are simplified to $\phi_j = [1,0,1,0]^\textsc{t}/\sqrt{\mathcal{N}}$, and $\overline{\phi}_j = [0,1,0,-1]^\textsc{t}/\sqrt{\mathcal{N}}$ when all adatoms are polarized along the $z$ axis, namely $\mathbf{e}^\mathbf{s}_j = \mathbf{e}_z$ for all $j$.

\section{Transformation of field operators}
\label{Appendix:Transformation}

In the previous section, in order to derive the low-energy $2N\times 2N$ effective Hamiltonian $\mathcal{H}_\mathrm{eff}$, we project the Nambu field operator $\hat{\Psi}(\bd r)$ on the Shiba basis $\vec{C} = [{c}_1, \dots, {c}_j, \dots, {c}_N, {c}^\dagger_1, \dots, {c}^\dagger_j, \dots, {c}^\dagger_N ]^\textsc{t}$:
\begin{align}
\hat{\Psi}(\bd r)
    = U_c \vec{C} \equiv [ \phi_1(\bd{r}), \dots, \phi_j(\bd{r}), \dots, \phi_N(\bd{r}), \overline{\phi}_1(\bd{r}), \dots, \overline{\phi}_j(\bd{r}), \dots, \overline{\phi}_N(\bd{r})]  \vec{C} =\sum_{j=1}^N \phi_j(\bd{r}){c}_j+\overline{\phi}_j(\bd{r}){c}^\dagger_j \,,
\label{Eq:TransformNambuToShiba}
\end{align}
where $\phi_j(\bd{r})$ is the YSR wavefunction over the entire space $\bd{r}$ induced by the impurity at site $j$. We omit the $S$ subscript of the Nambu field operator for simplicity. Based on Eqs.~\eqref{Eq:WaveFunctionOverSpace}-\eqref{Eq:ShibaBasis}, the YSR wavefunction reads
\begin{align}
    \phi_j(\bd{r})= -J_\textsc{e}(\bd{r}-\bd{r}_j) \tau_0\otimes(\mathbf{e}^\mathbf{s}_j \cdot \bd{\sigma}) \phi_j \,,\quad
    \phi_j \equiv \phi_j(\bd{r}_j)=[\cos (\zeta / 2), \mathrm{e}^{+\mathrm{i} \vartheta}\sin (\zeta / 2), \cos (\zeta / 2) , \mathrm{e}^{+\mathrm{i} \vartheta} \sin (\zeta / 2)]^\textsc{t} /\sqrt{\mathcal{N}} \,.
\label{Eq:ShibaWaveFunctionOverSpace}
\end{align}
A numerical diagonalization of $\mathcal{H}_\mathrm{eff}$ provides us with the positive eigenvalues $E_n$ and their eigenvectors $\varphi_n \equiv [p_n,q_n]^\textsc{t}$ in Eq.~\eqref{Eq:HeffTwoAngles}. The eigenvectors of the corresponding negative eigenvalues $-E_n$ are given by $\overline{\varphi}_n = \mathcal{C} \varphi_n = [q^*_n, p^*_n]^\textsc{t}$, where $\mathcal{C} = \tau_x \mathcal{K}$ is the particle-hole operator under the Shiba basis. Using these eigenvectors, the Shiba operators $\vec{C}$ can be expressed as low-energy Bogoliubons $\vec{D} = [{d}_0, \dots, {d}_n, \dots, {d}_{N-1}, {d}^\dagger_0, \dots, {d}^\dagger_n, \dots, {d}^\dagger_{N-1} ]^\textsc{t}$:
\begin{align}
    \vec{C} 
    = U_d \vec{D} \equiv 
    \begin{bmatrix}
        p_0 & \dots & p_n & \dots & p_{N-1} & q^*_0 & \dots & q^*_n & \dots & q^*_{N-1} \\
        q_0 & \dots & q_n & \dots & q_{N-1} & p^*_0 & \dots & p^*_n & \dots & p^*_{N-1}
    \end{bmatrix}
    \vec{D} \quad \Longleftrightarrow \quad {c}_j
    =\sum_{n=0}^{N-1} p_n(j) {d}_n+ q_n^*(j){d}^\dagger_n \,,
\label{Eq:TransformShibaToBogoliubon}
\end{align}
where we take the convention that the index of energy $n$ runs from $0$ to $N-1$, and ${d}_n$ annihilates a Bogoliubon of energy $E_n$. Combining these two transformations together, we express the original Nambu field operators in Bogoliubons:
\begin{align}
    \hat{\Psi}(\bd r)
    &= U_c U_d \vec{D}
    =\left[\dots, \sum_{j=1}^N \phi_j (\bd{r}) p_n(j) + \overline{\phi}_j (\bd{r}) q_n(j), \dots, \sum_{j=1}^N {\phi}_j (\bd{r}) q^*_n(j) + \overline{\phi}_j (\bd{r}) p^*_n(j), \dots\right] \vec{D} \nn \\
    &\equiv[\dots, \Phi_n(\bd{r}), \dots, \overline{\Phi}_n(\bd{r}), \dots]
    [{d}_0, \dots, {d}_n, \dots, {d}_{N-1}, {d}^\dagger_0, \dots, {d}^\dagger_n, \dots, {d}^\dagger_{N-1}]^\textsc{t} =\sum_{n=0}^{N-1} \Phi_n(\bd{r}){d}_n + \overline{\Phi}_n(\bd{r}){d}^\dagger_n \,.
\label{Eq:TransformNambuToBogoliubon}
\end{align}
For convenience we recast the last equation in a more commonly used form by defining
\begin{align}
\Phi_n(\bd{r})
    &= U_c \varphi_n
    = \sum_{j=1}^N \phi_j (\bd{r}) p_n(j) + \overline{\phi}_j (\bd{r}) q_n(j)
    \equiv[u_{n\uparrow}(\bd{r}),u_{n\downarrow}(\bd{r}),v_{n\downarrow}(\bd{r}),-v_{n\uparrow}(\bd{r})]^\textsc{t} \,.
    \label{Eq:WaveFunctionNambuToBogoliubon}
\end{align}
Taking out the component of $\hat{\Psi}(\bd r)$ in Eq.~\eqref{Eq:TransformNambuToBogoliubon}, we can express the time evolution of the electronic field by the quasi-particle:
\begin{align}
    \hat{\psi}_\sigma(\bd{r}, t) = \sum_{n=0}^{N-1} u_{n\sigma}(\bd{r}) {d}_n \mathrm{e}^{-\mathrm{i}E_nt} + v_{n\sigma}^*(\bd{r}) {d}^\dagger_n \mathrm{e}^{+\mathrm{i}E_nt}, \quad \sigma = \uparrow\downarrow \,.
    \label{Eq:TransformElectronToBogoliubon}
\end{align}

\section{Holstein-Primakoff transformation and the ferromagnetic lattice dynamics}
\label{Appendix:LatticeDynamics}

The magnetic Hamiltonian describing the  chain of impurities reads \cite{Kuster2022NonMajorana}
\begin{align}
    H_{\rm m}= \sum_{ij} J^\mathrm{ex}_{ij} \bd{S}_i\cdot \bd{S}_{j} - \sum_j \left(\frac{K_z}{2}(S_{j}^z)^2  - \gamma HS_{j}^z\right) + \sum_{ij}D_{ij}(\bd{S}_i\times \bd{S}_{j})_z\,,
\end{align}
where the last term represents the  Dzyaloshinskii-Moriya interaction along the $z$ axis with the coupling strength $D_{ij}$ between site $i$ and site $j$ (which generalizes Eq.~\eqref{Eq:Hamiltonian} in the main text by accounting for next-nearest-neighbor interactions). We assume the magnetic ground state corresponds to all spins being aligned along the $z$ direction ($J^\mathrm{ex}_{ij}<0$), and examine the magnonic fluctuations around it. To that end, we introduce the Holstein-Primakoff transformation ($\hbar=1$) \cite{Holstein1940Field}:
\begin{align}
    S_j^z=S-a_j^\dagger a_j\,, \quad
    S_j^+=\sqrt{2S}\sqrt{1-\frac{a_j^\dagger a_j}{2S}}a_j\approx\sqrt{2S}a_j\,, \quad
    S_j^-=\sqrt{2S}a_j^\dagger\sqrt{1-\frac{a_j^\dagger a_j}{2S}}\approx\sqrt{2S}a_j^\dagger\,,
\end{align}
where $a_j$ ($a_j^\dagger$) are the annihilation (raising) bosonic operators at position $j$ in the lattice, and we assumed that $\langle n_j\rangle \equiv \langle a_j^\dagger a_j\rangle\ll S$, where $\langle\dots\rangle$ is the expectation value over the (thermal) equilibrium magnetic state.  We can therefore express the above Hamiltonian in terms of the magnonic operators, keeping only terms up to quadratic order:
\begin{align}
    H_{\rm m}
    \approx\sum_{ij}J^\mathrm{ex}_{ij} S(a_i^\dagger a_j + a_j^\dagger a_i-n_i-n_j) + \sum_j (K_z S - \gamma H)n_j -\mathrm{i} \sum_{ij} D_{ij} S(a_i^\dagger a_j - a_j^\dagger a_i)\,.
    \label{Eq:MagnonicHamiltonian}
\end{align}
To gain a qualitative estimate of the magnonic spectrum, we first consider periodic boundary conditions and transform the Hamiltonian to the Fourier space:
\begin{align}
    H_{\rm m}&=\sum_k\epsilon_ka_k^\dagger a_k \,, \quad
    a_j = \frac{1}{\sqrt{N}}\sum_k \mathrm{e}^{+\mathrm{i}jk}a_k \,, \quad
    k=2\pi n/N\,, \quad n=0,\dots N-1 \,, \nn \\
    \epsilon_k &= 2 J^\mathrm{ex}_1 S [\cos (k)-1] + 2D_1 S\sin (k) + 2J^\mathrm{ex}_2 S [\cos (2k)-1] + 2D_2 S\sin (2k) + (K_zS-\gamma H )\,,
    \label{Eq:MagnonicSpectrum}
\end{align}
where $J^\mathrm{ex}_{1(2)}$ and $D_{1(2)}$ are the (next-)nearest-neighbor couplings, and we set the lattice spacing $a=1$. Assuming $K_zS>\gamma H $, which ensures that the ground state corresponds to the perpendicular alignment,  the lowest magnonic mode pertains to uniform precessions, $k=0$, with energy $\epsilon_{\rm m} \equiv K_zS-\gamma H $. The separation between this mode and the first excited one ($k=1$) is
\begin{align}
    \delta\epsilon_{10} \equiv
    \epsilon_1 - \epsilon_{\rm m} 
    = 2J^\mathrm{ex}_1 S [\cos (2 \pi /N) - 1] + 2D_1 S \sin (2 \pi /N) + 2J^\mathrm{ex}_2 S [\cos (4 \pi /N) - 1] + 2D_2 S \sin (4 \pi /N) \,.  
   \label{Eq:MagnonicSeparation}
\end{align}
Now we consider a more realistic case for an open chain. Through numerically diagonalizing Eq.~\eqref{Eq:MagnonicHamiltonian} with open boundary conditions, we show the magnonic spectrum $\epsilon_{n}$ and the separation $\delta\epsilon_{10}$ as a function of the number of impurities $N$ in Fig.~\ref{Fig:MagnonicSpectrum}. The specific parameters are adopted from a recent experiment in Ref.~\cite{Kuster2022NonMajorana} and can be found in Appendix~\ref{Appendix:NumericalDetails}. We find the uniform magnonic mode is well separated from the excited states when both the nearest-neighbor and next-nearest-neighbor couplings are accounted for. Hence, in the main text, we can only focus on the nearest-neighbor case and define $J_\mathrm{ex} \equiv J^\mathrm{ex}_1$, $D \equiv D_1$. The magnetic Hamiltonian in Eq.~\eqref{Eq:MagnonicHamiltonian} is reduced to 
\begin{align}
    H^{\rm NN}_{\rm m}
    = J_\mathrm{ex}S \sum_{j=1}^{N-1} ( a_j^\dagger a_{j+1} + a_{j+1}^\dagger a_j - n_j - n_{j+1} ) + \sum_{j=1}^N (K_z S - \gamma H ) n_j - \mathrm{i}D S \sum_{j=1}^{N-1} (a_j^\dagger a_{j+1} - a_{j+1}^\dagger a_j)\,.
    \label{Eq:MagnonicSpectrumOBC}
\end{align}
Using the properties of Toeplitz matrices, the eigenvalues of $\overline{H^{\rm NN}_{\rm m}} = H^{\rm NN}_{\rm m} - J_\mathrm{ex}S (n_1 + n_N)$ are exactly solvable as \cite{Gray2006Toeplitz}
\begin{align}
    \overline{\epsilon_k} = -2 S [\sqrt{J_\mathrm{ex}^2+D^2} \cos(k) + J_\mathrm{ex}] +(K_z S -\gamma  H) \,, \quad
    k= \pi n/(N+1)\,, \quad n = 1,\dots N \,. 
\end{align}
Since $J_\mathrm{ex} <0$, for $|J_\mathrm{ex}| \gg |D|$, the energy of the uniform mode ($k=0$) will be the same as periodic boundary conditions, $\overline{\epsilon_{\rm m}} = \epsilon_{\rm m} = K_z S -\gamma H$. The gap between this mode and the first excited one ($k=1$) is
\begin{align}
    \overline{\delta\epsilon_{10}} \equiv
    \overline{\epsilon_1} - \overline{\epsilon_{\rm m}} 
    = 2 S \sqrt{J_\mathrm{ex}^2+D^2} [\cos(\pi/(N+1)) - 1]  
    \approx \pi^2 S \sqrt{J_\mathrm{ex}^2+D^2} / (N+1)^2 \,,  
   \label{Eq:MagnonicSeparationLowerBound}
\end{align}
which gives a lower bound for the magnonic gap $\delta\epsilon_{10}$ of the Hamiltonian \eqref{Eq:MagnonicSpectrumOBC} with open boundary conditions. The level spacing between other low-energy modes can be estimated as $\overline{\epsilon_{n+1}} - \overline{\epsilon_{n}} = (2n+1) \pi^2 S \sqrt{J_\mathrm{ex}^2+D^2} / (N+1)^2$. We have numerically verified the asymptotic behavior of the above analytical results for low-energy modes up to $n=3$. Therefore, in the following discussions, we focus on the uniform mode as a single bosonic mode that can couple to the Majorana zero mode, rather than a bosonic bath, since its linewidth is smaller than the level spacing. 

The interaction Hamiltonian between the spins and the superconducting electrons reads
\begin{align}
    H_{\rm e-m}
    &= - J \sum_{j,\sigma,\sigma'} \psi^\dagger_{\sigma}(\bd{r}_j)(\bd{S}_j \cdot \boldsymbol{\sigma})_{\sigma\sigma'}\psi_{\sigma'}(\bd{r}_j)
    = - J \sum_{j,\sigma,\sigma'} \psi^\dagger_{\sigma}(\bd{r}_j)(S_j^- \sigma_+ + S_j^+ \sigma_- + S_j^z \sigma_z)_{\sigma\sigma'}\psi_{\sigma'}(\bd{r}_j)  \nn \\
    &\approx - J \sum_{j,\sigma,\sigma'} \psi^\dagger_{\sigma}(\bd{r}_j)\left[\sqrt{2S}(a_j^\dagger\sigma_+ + a_j\sigma_-) + (S-a_j^\dagger a_j) \sigma_z\right]_{\sigma\sigma'}\psi_{\sigma'}(\bd{r}_j) \,.
\end{align}
Next, projecting this Hamiltonian onto the uniform mode gives
\begin{align}
    H_{\rm e-m}
    &\approx \frac{J}{N} \sum_{j,\sigma,\sigma'}  \left\{ n_0 \psi^\dagger_{\sigma}(\bd{r}_j)(\sigma_z)_{\sigma\sigma'}\psi_{\sigma'}(\bd{r}_j) - \sqrt{2NS} [ a_0^\dagger \psi^\dagger_{\sigma}(\bd{r}_j) (\sigma_+)_{\sigma\sigma'}\psi_{\sigma'}(\bd{r}_j) + a_0 \psi^\dagger_{\sigma}(\bd{r}_j) (\sigma_-)_{\sigma\sigma'} \psi_{\sigma'}(\bd{r}_j) ] \right\} \nn \\
    &\equiv \frac{J}{N} [n_0 \Sigma_z - \sqrt{2NS}(a_0^\dagger \Sigma_+ + a_0 \Sigma_-) ]
    \equiv \frac{\Delta}{NS} [n_0 \widetilde{\Sigma}_z - \sqrt{2NS}(a_0^\dagger \widetilde{\Sigma}_+ + a_0 \widetilde{\Sigma}_-)]\,.
\end{align}
To elucidate the influence of the superconductor on the uniform mode, we employ the equation of motion method. This is appropriate if one plans to explore quantum effects. The equation of motion for the $k=0$ mode thus becomes
\begin{align}
    \dot{a}_0(t)
    &=\mathrm{i}[H_{\rm el} + H_{\rm m}+H_{\rm e-m}, a_0(t)]
    =-\mathrm{i} \epsilon_{\rm m} a_0(t) - \mathrm{i} \frac{\Delta}{NS} \left[a_0(t) \widetilde{\Sigma}_z(t) + \sqrt{2NS} a_0(t) \widetilde{\Sigma}_+(t)\right] \,,
    \label{Eq:OriginalEoM}
\end{align}
where all the operators are evolving in the Heisenberg picture. By introducing $U(t, t_0) = \mathcal{T} \exp[-\mathrm{i} \int_{t_0}^t \mathrm{d} t' H^\mathrm{I}_{\rm e-m}(t')]$, we express the spin operators in terms of their isolated evolution (the superscript I denotes operators in the interaction picture):
\begin{align}
    \widetilde{\Sigma}_{z}(t)
    &=U(t_0, t) \widetilde{\Sigma}^{\rm I}_{z}(t) U(t, t_0)
    \approx \widetilde{\Sigma}^{\rm I}_{z}(t) \,, \nn \\
    \widetilde{\Sigma}_+(t)
    &= U(t_0, t) \widetilde{\Sigma}^{\rm I}_{+}(t) U(t, t_0) 
    \approx \widetilde{\Sigma}^\mathrm{I}_{+}(t) + \mathrm{i}\int_{t_0}^t \mathrm{d} t'[H^\mathrm{I}_{\rm e-m}(t'),\widetilde{\Sigma}^\mathrm{I}_{+}(t)] \nn \\
    &=\widetilde{\Sigma}^{\rm I}_+(t)+ \mathrm{i} \frac{\Delta}{NS} \int_{t_0}^t \mathrm{d}t'[n_0\widetilde{\Sigma}^{\rm I}_z(t') - a_0^\dagger(t')\sqrt{2NS}\widetilde{\Sigma}^{\rm I}_+(t') - a_0(t')\sqrt{2NS}\widetilde{\Sigma}^{\rm I}_-(t'), \widetilde{\Sigma}^{\rm I}_+(t)] \nn \\
    &\approx \widetilde{\Sigma}^{\rm I}_+(t) -\frac{\Delta}{NS}  \sqrt{2NS} a_0 (t) \left\{ - \mathrm{i} \int_{-\infty}^\infty \mathrm{d}\tau \mathrm{e}^{\mathrm{i} \epsilon_{\rm m} \tau} \theta(\tau) [\widetilde{\Sigma}^{\rm I}_+(\tau), \widetilde{\Sigma}^{\rm I}_-(0)] \right\} \,,
\end{align}
where for $\widetilde{\Sigma}_{z}(t)$ we retained only the zeroth order, because of the factor $1/N$ in Eq.~\eqref{Eq:OriginalEoM}. Yet, we keep the first-order correction in $\widetilde{\Sigma}_{+}(t)$ since in the zeroth order $\langle\widetilde{\Sigma}^{\rm I}_+(t)\rangle \approx 0$, and the first order term contributes the same order $1/N$ to the following equation of motion. Specifically, the equation of motion in Eq.~\eqref{Eq:OriginalEoM} for the uniform magnon becomes
\begin{align}
    \dot{a}_0(t)
    &= - \mathrm{i} \epsilon_{\rm m} a_0(t) - \mathrm{i} \frac{\Delta}{NS}  a_0(t)\langle\widetilde{\Sigma}_z\rangle - \mathrm{i} \frac{2\Delta^2}{NS}  a_0(t)\Pi_{+-}(\epsilon_{\rm m})-\kappa_{\rm m}a_0(t)+h(t)\,, \nn \\
    \Pi_{+-}(\epsilon_{\rm m})&= - \mathrm{i} \int_{-\infty}^\infty \mathrm{d}\tau \mathrm{e}^{\mathrm{i} \epsilon_{\rm m} \tau} \theta(\tau) \langle[\widetilde{\Sigma}_+(\tau), \widetilde{\Sigma}_-(0)]\rangle\,,
    \label{Eq:SusceptibilityEoM}
\end{align}
where $\langle\dots\rangle$ represents the average over the electronic state and thus the superscript for the interaction picture can be omitted. We disregarded terms that do not contain the magnon degrees of freedom. Furthermore, we included a decay of the magnon mode which is quantified by the rate $\kappa_{\rm m}$, as well as a driving field $h(t)$ which triggers the dynamics of the magnon. The decay can be due to either Gilbert damping, or other mechanisms active for the impurity spins. This equation represents the analog of the input-output expression utilized in quantum optics and can be employed to quantify the magnonic field. Assuming $h(t)=h_0\mathrm{e}^{-\mathrm{i}\omega t}$, and switching with the magnons to the Fourier space $a_0(\omega) = \int \mathrm{d} t a_0(t) \mathrm{e}^{+\mathrm{i}\omega t}$, we find
\begin{align}
    a_0(\omega)
    =\frac{\mathrm{i} h_0}{\displaystyle\omega-\left[\epsilon_{\rm m}+ \frac{\Delta}{NS} \langle\widetilde{\Sigma}_z\rangle+ \frac{2\Delta^2}{NS} \Pi_{+-}(\epsilon_{\rm m}) \right]+\mathrm{i}\kappa_{\rm m}} \,.
\end{align}
Therefore, the magnon resonance frequency and its decay, respectively,  are shifted by 
\begin{align}
    \delta\epsilon_{\rm m}
    =\frac{\Delta}{NS} \langle\widetilde{\Sigma}_z\rangle + \frac{2\Delta^2}{NS} {\rm Re}\Pi_{+-}(\epsilon_{\rm m})\,,\quad
    \delta\kappa_{\rm m}
    =-\frac{2\Delta^2}{NS} {\rm Im}\Pi_{+-}(\epsilon_{\rm m}) \,.
\end{align}
These changes are analogous to electron-phonon interactions on Majorana zero modes in topological superconducting nanowires \cite{Aseev2019Degeneracy}, whose thermally excited phonons could conversely result in a broadening of Majorana zero modes with a linewidth $\gamma_\mathrm{ph}\propto\exp(-\Delta/T)$. For temperatures lower than the topological gap, $T\ll\Delta$, the phonon-induced decay is negligible and, by analogy, so is the magnon-induced decay, $\gamma_\mathrm{m}\propto\exp(-\Delta/T)$. 

In addition, the coherent magnonic driving $a_0 \equiv \alpha_0 + \tilde{a}_0$ could also excite the Majorana zero modes into the bulk states, where $\alpha_0=h_0/(\omega_0-\omega + \mathrm{i}\kappa_0)$ is the classical component with $\kappa_0$ being its decay, and $\tilde{a}_0$ captures the quantum fluctuation. The Majorana decay is dominated by the classical coherent driving field $\alpha_0$ with the Hamiltonian
\begin{align}
    H_{\rm drive}(t) 
    = -\sqrt{\frac{2 \Delta^2}{NS}} \left(\alpha_0^* \mathrm{e}^{-\mathrm{i}\omega t} \widetilde{\Sigma}_+ +\alpha_0 \mathrm{e}^{+\mathrm{i}\omega t}\widetilde{\Sigma}_- \right) \,.
\end{align}
By the time-dependent perturbation theory, the probability to excite the ground state of parity $\mathcal{P}$ to a bulk state $n$ reads,
\begin{align}
    P_{\mathcal{P}\rightarrow n} 
    = \frac{2|\alpha_0|^2\Delta^2}{NS\hbar^2} \left | \bra{n} \widetilde{\Sigma}_{+} \ket{\mathcal{P}} \frac{\sin[(E_n - E_\mathcal{P} - \omega) t/2]}{(E_n - E_\mathcal{P} - \omega)/2} \right |^2 
    \xrightarrow{\omega = E_n - E_\mathcal{P}} \frac{2|\alpha_0|^2\Delta^2}{NS\hbar^2} | \bra{n} \widetilde{\Sigma}_{+} \ket{\mathcal{P}} |^2  t^2 \,,
\end{align}
where $|\alpha_0|^2$ is the average number of magnons determined by the input power onto the microwave drive, and $\bra{n} \widetilde{\Sigma}_{+} \ket{0} = \overline{B}_{n0}(\sigma_+)/2$ and $\bra{n} \widetilde{\Sigma}_{+} \ket{1} = A_{n0}(\sigma_+)$ can be calculated using Eqs.~\eqref{Eq:TotalSpinABC}-\eqref{Eq:MatricesABC} in the next section. To prevent the lowest states from being depopulated, the driving needs to be performed within the time scale $t_0$ to ensure $P_{\mathcal{P}\rightarrow n}\ll1$: 
\begin{align}
    t &\ll t_0 \equiv \frac{\sqrt{NS}}{|\alpha_0| \, |\bra{n} \widetilde{\Sigma}_{+} \ket{\mathcal{P}}|} \frac{\hbar}{\Delta} \,,
\end{align} 

\begin{figure}[t]
    \centering
    \begin{minipage}{0.49\textwidth}
        \centering
        \includegraphics[width=\textwidth]{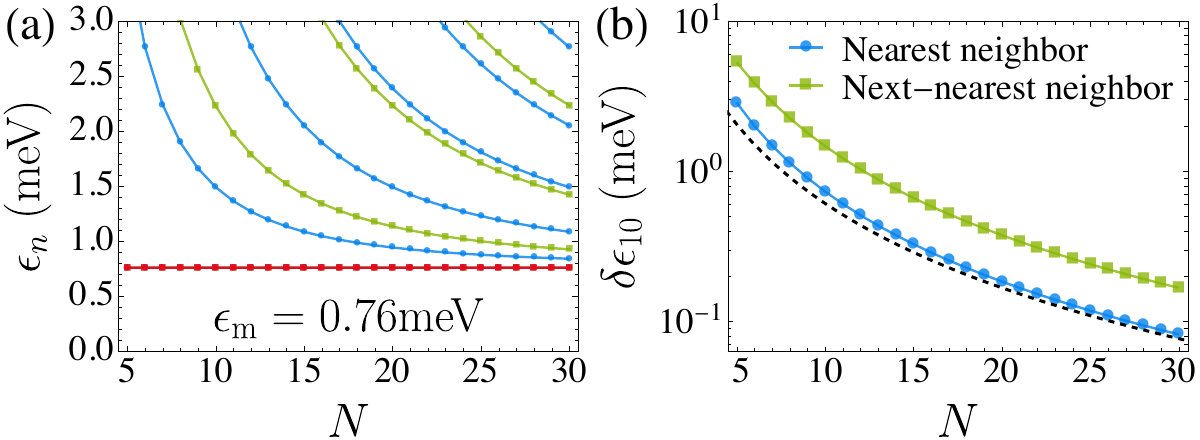}
        \caption{The magnonic spectra vary as the number of spins $N$ in the chain with open boundary conditions. In panel (a), the blue (green) lines are the full spectrum $\epsilon_n$  with (next-)nearest-neighbor couplings accounted for. The uniform mode energy (red) $\epsilon_{\rm m}=0.76$ meV at $H=0$, which can be lowered by switching on a finite $H$. Panel (b) shows the separation $\delta\epsilon_{10}$ between $\epsilon_{\rm m}$ and the first excited mode, whose lower bound [Eq.~\eqref{Eq:MagnonicSeparationLowerBound}] is indicated by the dashed line. The parameters are adopted from Ref.~\cite{Kuster2022NonMajorana}, where Cr magnetic adatoms were deposited on the top of the Nb superconducting substrate.}
        \label{Fig:MagnonicSpectrum}
    \end{minipage}\hfill
    \begin{minipage}{0.49\textwidth}
        \centering
        \includegraphics[width=\textwidth]{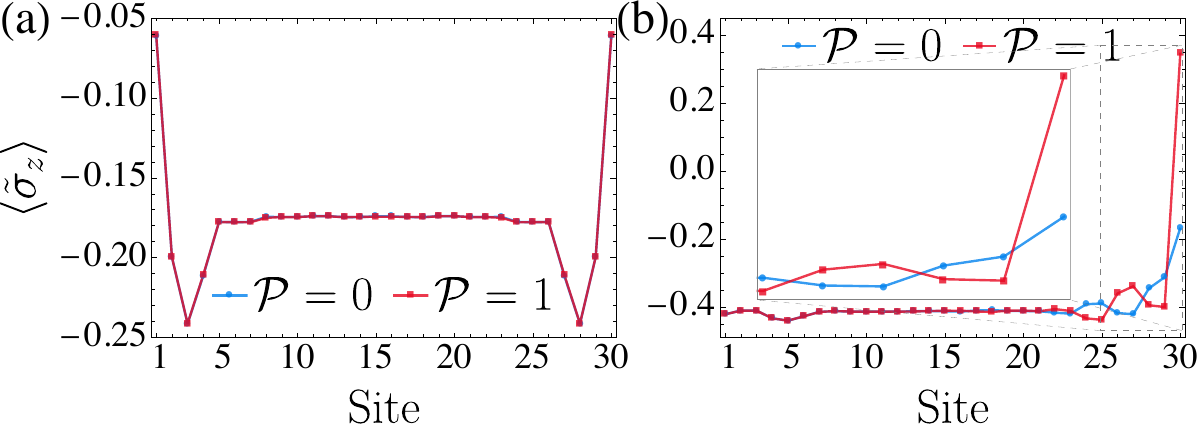}
        \caption{Majorana zero modes vs. trivial zero modes in local spin expectation values. Panel (a) and (b) are the site-dependent spin expectation values $\langle\tilde{\sigma}_z\rangle$ along the $z$-axis for $N=30$ in the topological phase and the trivial phase, respectively. All other parameters are set the same as Fig.~\ref{Fig:TotalSpin} in the main text. The blue (red) lines refer to the $\mathcal{P}=0(1)$ parity. As the trivial zero mode is obtained by fine tuning the last site, $\langle\tilde{\sigma}_z\rangle$ in (b) shows a strong deviation between two parities at the end of the chain.}
        \label{Fig:SiteSpin}
    \end{minipage}
\end{figure}

\section{Spin susceptibility and visibility}
\label{Appendix:Susceptibility}

To calculate the spin susceptibility in Eq.~\eqref{Eq:SusceptibilityEoM}, we need to write down the time evolution of the spin operator $\widetilde{\Sigma}_\pm(t)$, which can be simply obtained by renormalizing the wavefunction of $\Sigma_\pm(t)$ (see Appendix~\ref{Appendix:NumericalDetails} for details). Hence in this section we only focus on the calculation of $\Sigma_\pm(t)$. By use of Eq.~\eqref{Eq:TransformElectronToBogoliubon}, we can express $\Sigma_\pm(t)$ with a set of Bogoliubons:
\begin{align}
\label{Eq:TotalSpinABC}
\Sigma_\pm(t)
    &= \frac{1}{2} \sum_j \hat{\Psi}^\dagger(\bd{r}_j, t) (\tau_0\otimes \sigma_\pm)\hat{\Psi}(\bd{r}_j, t) 
    = \sum_{j,\sigma,\sigma'} \hat{\psi}^\dagger_\sigma(\bd{r}_j, t) (\sigma_\pm)_{\sigma\sigma'} \hat{\psi}_{\sigma'}(\bd{r}_j, t) \\
    &= \sum_{j,\sigma,\sigma',n,m}\left[u^*_{n\sigma}(\bd{r}_j){d}^\dagger_n \mathrm{e}^{ + \mathrm{i} E_nt}+v_{n\sigma}(\bd{r}_j){d}_n \mathrm{e}^{-\mathrm{i}E_nt}\right] (\sigma_\pm)_{\sigma\sigma'} \left[u_{m\sigma'}(\bd{r}_j){d}_m \mathrm{e}^{-\mathrm{i} E_mt} + v^*_{m\sigma'}(\bd{r}_j){d}^\dagger_m \mathrm{e}^{\mathrm{i}E_mt}\right] \nn \\
    &= \sum_{j,\sigma,\sigma',n,m} \Big\{\left[u^*_{n\sigma}(\bd{r}_j)(\sigma_\pm)_{\sigma\sigma'}u_{m\sigma'}(\bd{r}_j)-v_{m\sigma}(\bd{r}_j)(\sigma_\pm)_{\sigma\sigma'}v^*_{n\sigma'}(\bd{r}_j)\right] {d}^\dagger_n{d}_m \mathrm{e}^{+\mathrm{i}(E_n-E_m)t} \nn \\
    &\qquad \qquad \quad \; +\frac{v_{n\sigma}(\bd{r}_j)(\sigma_\pm)_{\sigma\sigma'}u_{m\sigma'}(\bd{r}_j)-v_{m\sigma}(\bd{r}_j)(\sigma_\pm)_{\sigma\sigma'}u_{n\sigma'}(\bd{r}_j)}{2}{d}_n{d}_m \mathrm{e}^{-\mathrm{i}(E_n+E_m)t}\nn\\
    &\qquad \qquad \quad \; +\frac{u^*_{n\sigma}(\bd{r}_j)(\sigma_\pm)_{\sigma\sigma'}v^*_{m\sigma'}(\bd{r}_j)-u^*_{m\sigma}(\bd{r}_j)(\sigma_\pm)_{\sigma\sigma'}v^*_{n\sigma'}(\bd{r}_j)}{2}{d}^\dagger_n{d}^\dagger_m  \mathrm{e}^{+\mathrm{i}(E_n+E_m)t} + \delta_{nm}v_{m\sigma}(\sigma_\pm)_{\sigma\sigma'}v^*_{n\sigma'} \Big\} \nn\\
    &\equiv \sum_{n,m} \Big[ A_{nm}(\sigma_\pm) {d}^\dagger_n{d}_m \mathrm{e}^{+\mathrm{i}(E_n-E_m)t} + \frac{B_{nm}(\sigma_\pm)}{2} {d}_n{d}_m \mathrm{e}^{-\mathrm{i}(E_n+E_m)t} + \frac{\overline{B}_{nm}(\sigma_\pm)}{2} {d}^\dagger_n{d}^\dagger_m  \mathrm{e}^{+\mathrm{i}(E_n+E_m)t} + C_{nm}(\sigma_\pm)\Big] \nn \,.
\end{align}
One can find the above matrices obey $A(\sigma_\pm) = A(\sigma_\pm^\dagger)^\dagger, B(\sigma_\pm) = -B(\sigma_\pm)^\textsc{t}, \overline{B}(\sigma_\pm) = B(\sigma_\pm^\dagger)^\dagger$, or more explicitly,
\begin{align}
A_{nm}(\sigma_\pm)
    &= \sum_{j,\sigma,\sigma'} [u^*_{n\sigma}(\bd{r}_j)(\sigma_\pm)_{\sigma\sigma'}u_{m\sigma'}(\bd{r}_j)-v_{m\sigma}(\bd{r}_j)(\sigma_\pm)_{\sigma\sigma'}v^*_{n\sigma'}(\bd{r}_j)]
    = \sum_j \Phi_n^\dagger(\bd{r}_j) \sigma_\pm \Phi_m(\bd{r}_j) = A_{mn}^*(\sigma_\pm^\dagger) \,, \nn \\
B_{nm}(\sigma_\pm)
    &= \sum_{j,\sigma,\sigma'} [v_{n\sigma}(\bd{r}_j)(\sigma_\pm)_{\sigma\sigma'}u_{m\sigma'}(\bd{r}_j)-v_{m\sigma}(\bd{r}_j)(\sigma_\pm)_{\sigma\sigma'}u_{n\sigma'}(\bd{r}_j)]
    = \sum_j \overline{\Phi}_n^\dagger(\bd{r}_j) \sigma_\pm \Phi_m(\bd{r}_j) = -B_{mn}(\sigma_\pm)\,, \nn \\
\overline{B}_{nm}(\sigma_\pm)
    &= \sum_{j,\sigma,\sigma'} [u^*_{n\sigma}(\bd{r}_j)(\sigma_\pm)_{\sigma\sigma'}v^*_{m\sigma'}(\bd{r}_j)-u^*_{m\sigma}(\bd{r}_j)(\sigma_\pm)_{\sigma\sigma'}v^*_{n\sigma'}(\bd{r}_j)]
    = \sum_j \Phi_n^\dagger(\bd{r}_j) \sigma_\pm \overline{\Phi}_m(\bd{r}_j) = B_{mn}^*(\sigma_\pm^\dagger) \,.
    \label{Eq:MatricesABC}
\end{align}
With these matrices, now we can write the spin-susceptibility $\Pi_{+-}(t) = -\mathrm{i} \theta(t) \expval{[\Sigma_+(t), \Sigma_-(0)]}$ in a compact form:
\begin{align}
    \Pi_{+-}(t)
    =-\mathrm{i} \theta(t) \sum_{n,m,p,q} \Big\langle \Big[ 
    &A_{nm}(\sigma_+){d}^\dagger_n {d}_m \mathrm{e}^{+\mathrm{i}(E_n-E_m)t} + \frac{B_{nm}(\sigma_+)}{2} {d}_n {d}_m \mathrm{e}^{-\mathrm{i}(E_n+E_m)t} + \frac{\overline{B}_{nm}(\sigma_+)}{2} {d}^\dagger_n {d}^\dagger_m \mathrm{e}^{+\mathrm{i}(E_n+E_m)t}, \nn \\
    &A_{pq}(\sigma_-){d}^\dagger_p {d}_q + \frac{B_{pq}(\sigma_-)}{2} {d}_p {d}_q + \frac{\overline{B}_{pq}(\sigma_-)}{2} {d}^\dagger_p {d}^\dagger_q \Big] \Big\rangle \,.
\end{align}
Note that $C_{nm}(\sigma_\pm)$ in the last line of Eq.~\eqref{Eq:TotalSpinABC} is a C-number and thus does not contribute to the above commutator. As the number of Bogoliubons is conserved, the previous equation can be further simplified:
\begin{align}
    \Pi_{+-}(t)=-\mathrm{i} \theta(t) \sum_{n,m,p,q} \Big\langle 
    &A_{nm}(\sigma_+)A_{pq}(\sigma_-)[{d}^\dagger_n {d}_m,{d}^\dagger_p {d}_q ] \mathrm{e}^{+\mathrm{i}(E_n-E_m)t} 
    + \frac{B_{nm}(\sigma_+)\overline{B}_{pq}(\sigma_-)}{4} [ {d}_n {d}_m,{d}^\dagger_p {d}^\dagger_q ]\mathrm{e}^{-\mathrm{i}(E_n+E_m)t} \nn \\
    + &\frac{\overline{B}_{nm}(\sigma_+)B_{pq}(\sigma_-)}{4} [{d}^\dagger_n {d}^\dagger_m , {d}_p {d}_q ] \mathrm{e}^{+\mathrm{i}(E_n+E_m)t} \Big\rangle \,.
\end{align}
Next, we use the identity, $[AB, CD]=A[B,CD]+[A,CD]B=A(\{B,C\}D-C\{B,D\})+(\{A,C\}D-C\{A,D\})B$, and $\langle a^\dagger_na_m\rangle=\delta_{nm}f_n$ with $f_n$ the occupation number 
at energy $E_n$. We obtain
\begin{align}
    \Pi_{+-}(t)=-\mathrm{i} \theta(t) \sum_{n,m} \Big [
    &A_{nm}(\sigma_+)A_{mn}(\sigma_-) (f_n - f_m) \mathrm{e}^{+\mathrm{i}(E_n-E_m)t} + \frac{B_{nm}(\sigma_+)\overline{B}_{nm}(\sigma_-)}{2} (f_m + f_n - 1)\mathrm{e}^{-\mathrm{i}(E_n+E_m)t} \nn\\
    +&\frac{\overline{B}_{nm}(\sigma_+)B_{nm}(\sigma_-)}{2} (1 - f_m - f_n) \mathrm{e}^{+\mathrm{i}(E_n+E_m)t} \Big ] \,.
\end{align}
Finally, we use the relation $\overline{B}_{nm}(\sigma_\pm) = B_{mn}^*(\sigma_\pm^\dagger) = B_{mn}^*(\sigma_\mp)$ and then transform the spin susceptibility into frequency space: 
\begin{align}
\int_{-\infty}^\infty \mathrm{d} t \mathrm{e}^{\mathrm{i} \omega t} \Pi_{+-}(t)
    =\sum_{n,m} \Big [
    &A_{nm}(\sigma_+)A_{mn}(\sigma_-)\frac{f_n-f_m}{\omega+E_n-E_m+\mathrm{i}\eta}
    +\frac{B_{nm}(\sigma_+) B_{mn}^*(\sigma_+)}{2} \frac{f_m+f_n-1}{\omega-E_n-E_m+\mathrm{i}\eta} \nn \\
    +&\frac{B_{mn}^*(\sigma_-) B_{nm}(\sigma_-)}{2} \frac{1-f_m-f_n}{\omega+E_n+E_m+\mathrm{i}\eta} \Big ]
    \equiv \Pi_{+-}(\omega) \,,
    \label{Eq:SusceptibilityFn}
\end{align}
where $\eta$ is an infinitesimal positive constant. Our goal is to study the role of the parity in the spin susceptibility. To this end, we focus on the zero-temperature limit and assume that the lowest-energy Bogoliubon ${d}_0$ has an exponentially small energy $E_0 \rightarrow 0$ and the gap $\Delta_\mathrm{eff} = E_1 - E_0$ is nonzero. Hence, there are two degenerate many-body states with different fermionic parities: $\ket{\text{vac}}$ and ${d}_0^\dagger\ket{\text{vac}}$. Under this circumstance, the occupation number $f_n=0$ for $n\geq1$ and $\mathcal{P}\equiv f_0=0,1$ which depends on the parity of the many-body state. Under the aforementioned assumptions, the spin susceptibility reads 
\begin{align}
\Pi_{+-} (\omega, \mathcal{P})
    =\sum_{n>0} &\left[
    \frac{\mathcal{P}A_{0n}(\sigma_+)A_{n0}(\sigma_-)}{\omega-E_n+E_0+\mathrm{i}\eta} 
    - \frac{(1-\mathcal{P})B_{0n}(\sigma_+) B_{n0}^*(\sigma_+) }{\omega-E_n-E_0+\mathrm{i}\eta}
    - \frac{\mathcal{P}A_{n0}(\sigma_+)A_{0n}(\sigma_-)}{\omega+E_n-E_0+\mathrm{i}\eta} 
    + \frac{(1-\mathcal{P})B_{n0}^*(\sigma_-) B_{0n}(\sigma_-)}{\omega+E_n+E_0+\mathrm{i}\eta} \right]\nn \\
    +\sum_{n,m>0} &\left[
    \frac{B_{mn}^*(\sigma_-) B_{nm}(\sigma_-)/2}{\omega+E_n+E_m+\mathrm{i}\eta}
	-\frac{B_{nm}(\sigma_+) B_{mn}^*(\sigma_+)/2}{\omega-E_n-E_m+\mathrm{i}\eta} \right] 
    \,.
    \label{Eq:SusceptibilityFull}
\end{align}
The last line represents the bulk-bulk contribution, which is independent of the parity. We use Eq.~\eqref{Eq:SusceptibilityFull} to calculate the spin susceptibility and its visibility for all figures. When $\Delta_\mathrm{eff}< \omega <2\Delta_\mathrm{eff}$, only the first two terms dominate the susceptibility, which can be written in a compact form: 
\begin{align}
    \Pi_{+-} (\omega, \mathcal{P})
    \approx \sum_{n>0} \left[
        \frac{\mathcal{P}A_{0n}(\sigma_+)A_{n0}(\sigma_-)}{\omega-E_n+E_0+\mathrm{i}\eta} 
        - \frac{(1-\mathcal{P})B_{0n}(\sigma_+) B_{n0}^*(\sigma_+) }{\omega-E_n-E_0+\mathrm{i}\eta} \right]
    = \sum_{n>0}\frac{-(-1)^\mathcal{P}\mathcal{O}^{\mathcal{P}+}_{0n}\mathcal{O}^{\mathcal{P}-}_{n0}}{\omega-E_n-(-1)^\mathcal{P} E_0 + \mathrm{i}\eta}\,, 
    \label{Eq:SusceptibilityParity}
\end{align}
where we define $\mathcal{O}^{1\pm}_{nm} \equiv A_{nm}(\sigma_\pm)$, $\mathcal{O}^{0+}_{nm} \equiv B_{nm}(\sigma_+)$, and $\mathcal{O}^{0-}_{nm} \equiv B_{nm}^*(\sigma_+) = \overline{B}_{mn}(\sigma_-)$. To sum up, we have
\begin{align}
    \mathcal{O}^{\mathcal{P}\pm}_{nm}
    = \sum_j[ \Phi_n^\dagger(\bd{r}_j)\delta_{\mathcal{P}1}+\overline{\Phi}_n^\dagger(\bd{r}_j)\delta_{\mathcal{P}0}] \sigma_\pm \Phi_m(\bd{r}_j) \,, \quad \text{except for} \quad 
    \mathcal{O}^{0-}_{nm} \equiv (\mathcal{O}^{0+}_{nm})^* \,.
\end{align}
By use of the identity $\lim_{\eta\rightarrow0^{+}} \frac{1}{x \pm \mathrm{i} \eta} = \mathrm{p.v.} (\frac{1}{x}) \mp \mathrm{i} \pi \delta(x)$, we find
\begin{align}
\Im \Pi_{+-} (\omega, \mathcal{P})
    &= \pi \sum_{n>0} (-1)^\mathcal{P}\mathcal{O}^{\mathcal{P}+}_{0n}\mathcal{O}^{\mathcal{P}-}_{n0} \delta \left(\omega-E_n-(-1)^\mathcal{P} E_0\right) \,,
\end{align}
and the visibility of the spin susceptibility is reduced to
\begin{align}
\mathcal{V}(\omega) 
    \equiv \frac{\Im\Pi_{+-}(\omega,0)-\Im\Pi_{+-}(\omega,1)}{\Im\Pi_{+-}(\omega,0)+\Im\Pi_{+-}(\omega,1)}
    = \frac{\mathcal{O}^{0+}_{0n}\mathcal{O}^{0-}_{n0} + \mathcal{O}^{1+}_{0n}\mathcal{O}^{1-}_{n0}}{\mathcal{O}^{0+}_{0n}\mathcal{O}^{0-}_{n0} - \mathcal{O}^{1+}_{0n}\mathcal{O}^{1-}_{n0}} \,.
    \label{Eq:VisibilitySM}
\end{align}

\section{Inversion symmetry and quantized visibility}
\label{Appendix:QuantizedVisibility}

When all the parameters are set uniform in Eq.~\eqref{Eq:HeffTwoAngles}, $\mathcal{H}_\mathrm{eff}$ has an inversion symmetry $[\mathcal{H}_\mathrm{eff}, \mathcal{S}] = 0$, where $\mathcal{S} = \tau_z \otimes \mathcal{I}$ is the inversion operator and $\mathcal{I}_{ij} = \delta_{i,N+1-j}$ that maps site $j$ into $N+1-j$. Hence the eigenvector $\varphi_n$ of $\mathcal{H}_\mathrm{eff}$ with energy $E_n$ is also an eigenvector of $\mathcal{S}$ with eigenvalue $S_n=\pm1$. To scrutinize the effect of the inversion symmetry, we first rewrite the matrices in Eq.~\eqref{Eq:MatricesABC} by use of $\Phi_n(\bd{r}_k) = U_c (\bd{r}_k) \varphi_n$ in Eq.~\eqref{Eq:WaveFunctionNambuToBogoliubon}:
\begin{align}
A_{nm}(\sigma_\pm)
    &= \sum_k \Phi_n^\dagger(\bd{r}_k) \sigma_\pm \Phi_m(\bd{r}_k)
    = \varphi_n^\dagger \left[\sum_k U^\dagger_c(\bd{r}_k) \sigma_\pm U_c(\bd{r}_k)\right] \varphi_m
    \equiv \varphi_n^\dagger X(\sigma_\pm) \varphi_m \,,  
X(\sigma_\pm) = 
    \begin{bmatrix}
        X_{11}(\sigma_\pm) & X_{12}(\sigma_\pm) \\
        X_{21}(\sigma_\pm) & X_{22}(\sigma_\pm)
    \end{bmatrix}\,, \nn \\
B_{nm}(\sigma_\pm)
    &= \sum_k \overline{\Phi}_n^\dagger(\bd{r}_k) \sigma_\pm \Phi_m(\bd{r}_k)
    = \overline{\varphi}_n^\dagger \left[\sum_k U^\dagger_c(\bd{r}_k) \sigma_\pm U_c(\bd{r}_k)\right] \varphi_m
    \equiv \overline{\varphi}_n^\dagger X(\sigma_\pm) \varphi_m \,.
\end{align}
By examining the components of $U_c(\bd{r}_k)$ in Eq.~\eqref{Eq:TransformNambuToShiba}, one can find the block of the $X(\sigma_\pm)$ matrix has the following properties,
\begin{align}
    X^{22}_{ij}(\sigma_\pm)
    &= \sum_{k} \overline{\phi}^\dagger_i (\bd{r}_k) \sigma_\pm \overline{\phi}_j(\bd{r}_k)
    = \sum_{k} \phi^\textsc{t}_i (\bd{r}_k) \sigma_y \sigma_\pm \sigma_y \phi^*_j(\bd{r}_k)
    = -\sum_{k} \phi^\textsc{t}_i (\bd{r}_k) \sigma_\pm^\textsc{t} \phi^*_j(\bd{r}_k)
    = -X^{11}_{ji}(\sigma_\pm) \,, \nn \\
    X^{21}_{ij}(\sigma_\pm)
    &= \sum_{k} \overline{\phi}^\dagger_i (\bd{r}_k) \sigma_\pm \phi_j(\bd{r}_k)
    = -\sum_{k} \phi^\textsc{t}_i (\bd{r}_k) \sigma_\pm^\textsc{t} \tau_y \sigma_y \phi_j(\bd{r}_k)
    = -\sum_{k} \phi^\textsc{t}_i(\bd{r}_k)  \sigma_\pm^\textsc{t} \overline{\phi}^*_j (\bd{r}_k)
    = -X^{21}_{ji}(\sigma_\pm) \,, 
\end{align}
which mean $X_{22}(\sigma_\pm) = X_{22}^\dagger(\sigma_\pm^\dagger) = - X_{11}^\textsc{t}(\sigma_\pm) = - X_{11}^*(\sigma_\pm^\dagger) $ and $X_{21}(\sigma_\pm) = -X_{21}^\textsc{t}(\sigma_\pm) = X_{12}^\dagger(\sigma_\pm^\dagger)$. The above properties are applicable for a general system which is not required to hold the inversion symmetry. 
If the system has an inversion symmetry, the Shiba spinor follows $\phi_{N+1-j} = \phi_{j}$ and thus one can find
\begin{align}
    \phi_{N+1-j}(\bd{r}_{k})
    &= J_\textsc{e}(\bd{r}_{k}-\bd{r}_{N+1-j}) \phi_{N+1-j} 
    = J_\textsc{e}[-(\bd{r}_{N+1-k}-\bd{r}_{j})] \phi_{j} 
    = \sigma_z J_\textsc{e}(\bd{r}_{N+1-k}-\bd{r}_{j}) \phi_{j}
    = \sigma_z \phi_{j}(\bd{r}_{N+1-k}) \,,
\end{align}
where we use the properties of integrals [Eq.~\eqref{Eq:IntegralExact}] inside $J_\textsc{e}(\bd{r})$ [Eq.~\eqref{Eq:Jmatrix}] that $I^\pm_{1,3}$ are even functions, whereas $I^\pm_{2,4}$ are odd functions. By using the above relation, the blocks of the $X(\sigma_\pm)$ matrix have additional properties:
\begin{align}
    X^{11}_{N+1-i,N+1-j}(\sigma_\pm)
    &= \sum_{k} \phi^\dagger_{N+1-i} (\bd{r}_k) \sigma_\pm \phi_{N+1-j}(\bd{r}_k)
    = \sum_{k} \phi^\dagger_i (\bd{r}_k) \sigma_z \sigma_\pm \sigma_z \phi_j(\bd{r}_k)
    = +X^{11}_{ij}(\sigma_z \sigma_\pm \sigma_z) \,,  \nn \\
    X^{12}_{N+1-i,N+1-j}(\sigma_\pm)
    &= \sum_{k} \phi^\dagger_i (\bd{r}_k) \sigma_z \sigma_\pm \tau_y \sigma_y  \sigma_z \phi^*_j(\bd{r}_k)
    = -\sum_{k} \phi^\dagger_i (\bd{r}_k) \sigma_z \sigma_\pm \sigma_z \overline{\phi}_j (r_k)
    = -X^{12}_{ij}(\sigma_z \sigma_\pm \sigma_z) \,.
\end{align}
Consequently, the $X(\sigma_\pm)$ matrix is transformed by the inversion symmetry operator $\mathcal{S}$ as follows:
\begin{align}
    \mathcal{S} X(\sigma_\pm) \mathcal{S}=
    \begin{bmatrix}
        \mathcal{I} X_{11}(\sigma_\pm) \mathcal{I} & -\mathcal{I} X_{12}(\sigma_\pm) \mathcal{I} \\
        -\mathcal{I} X_{21}(\sigma_\pm) \mathcal{I} & \mathcal{I} X_{22}(\sigma_\pm) \mathcal{I}
    \end{bmatrix}=
    \begin{bmatrix}
        X_{11}(\sigma_z \sigma_\pm \sigma_z) & X_{12}(\sigma_z \sigma_\pm \sigma_z) \\
        X_{21}(\sigma_z \sigma_\pm \sigma_z) & X_{22}(\sigma_z \sigma_\pm \sigma_z) 
    \end{bmatrix}
    =X(\sigma_z \sigma_\pm \sigma_z)
    =-X(\sigma_\pm) \,.
\end{align}
This has important consequences for the matrices used for the susceptibility $\Pi_{+-}(\omega, \mathcal{P})$ in Eq.~\eqref{Eq:SusceptibilityFull}: 
\begin{align*}
    A_{nm}(\sigma_\pm)
        &= \varphi^\dagger_n X(\sigma_\pm) \varphi_m
        = \varphi^\dagger_n \mathcal{S} \mathcal{S} X(\sigma_\pm) \mathcal{S} \mathcal{S} \varphi_m 
        = -\varphi^\dagger_n \mathcal{S} X(\sigma_\pm) \mathcal{S} \varphi_m 
        = - S_n S_m \varphi^\dagger_n X(\sigma_\pm) \varphi_m 
        = - S_n S_m A_{nm}(\sigma_\pm)  \,, \\
    B_{nm}(\sigma_\pm)
        &= \overline{\varphi}^\dagger_n X(\sigma_\pm) \varphi_m 
        = \overline{\varphi}^\dagger_n \mathcal{S} \mathcal{S} X(\sigma_\pm) \mathcal{S} \mathcal{S} \varphi_m 
        = -\overline{\varphi}^\dagger_n \mathcal{S} X(\sigma_\pm) \mathcal{S} \varphi_m 
        = - \overline{S}_n S_m \overline{\varphi}^\dagger_n X(\sigma_\pm) \varphi_m
        = + S_n S_m B_{nm}(\sigma_\pm)  \,. 
\end{align*}
It directly implies that $A_{nm}(\sigma_\pm) B_{nm}(\sigma_\pm) = 0$. Using the notation defined in Eq.~\eqref{Eq:SusceptibilityParity}, we find $\mathcal{O}^{0\pm}_{nm}=0$ if $S_n \neq S_m$, while $\mathcal{O}^{1\pm}_{nm}=0$ if $S_n = S_m$. In other words, $\mathcal{O}^{0\pm}_{0n}$ is only nonzero for states that satisfy $S_0S_n=1$, while only states with $S_0S_n=-1$ lead to nonzero $\mathcal{O}^{1\pm}_{0n}$. Hence, the visibility in Eq.~\eqref{Eq:VisibilitySM} is reduced to $\mathcal{V}(\omega_n) = S_0 S_n$, which leads to the $\pm 1$ oscillations in Fig.~\ref{Fig:Susceptibility}.

\section{More numerical results for spin expectation value and visibility}
\label{Appendix:SpinExpectationValue}

\begin{figure}[t]
    \centering
    \begin{minipage}{0.49\textwidth}
        \centering
        \includegraphics[width=0.97\textwidth]{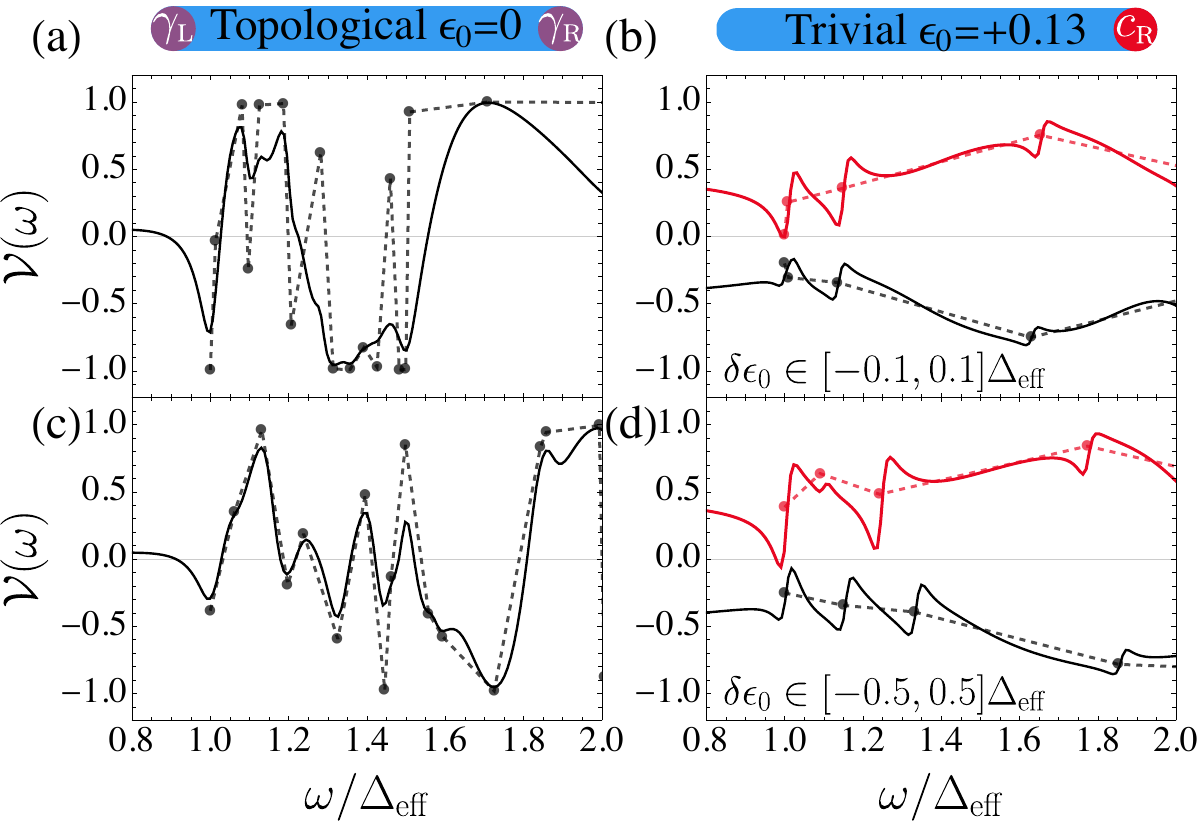}
        \caption{Specific realizations (without average) of the frequency dependence of the visibility $\mathcal{V}(\omega)$ in the presence of random onsite disorder $\delta\epsilon_0$. All parameters are set the same as Fig.~\ref{Fig:Disorders} in the main text.}
        \label{Fig:DisordersSupp}
    \end{minipage}\hfill
    \begin{minipage}{0.49\textwidth}
        \centering
        \includegraphics[width=0.98\textwidth]{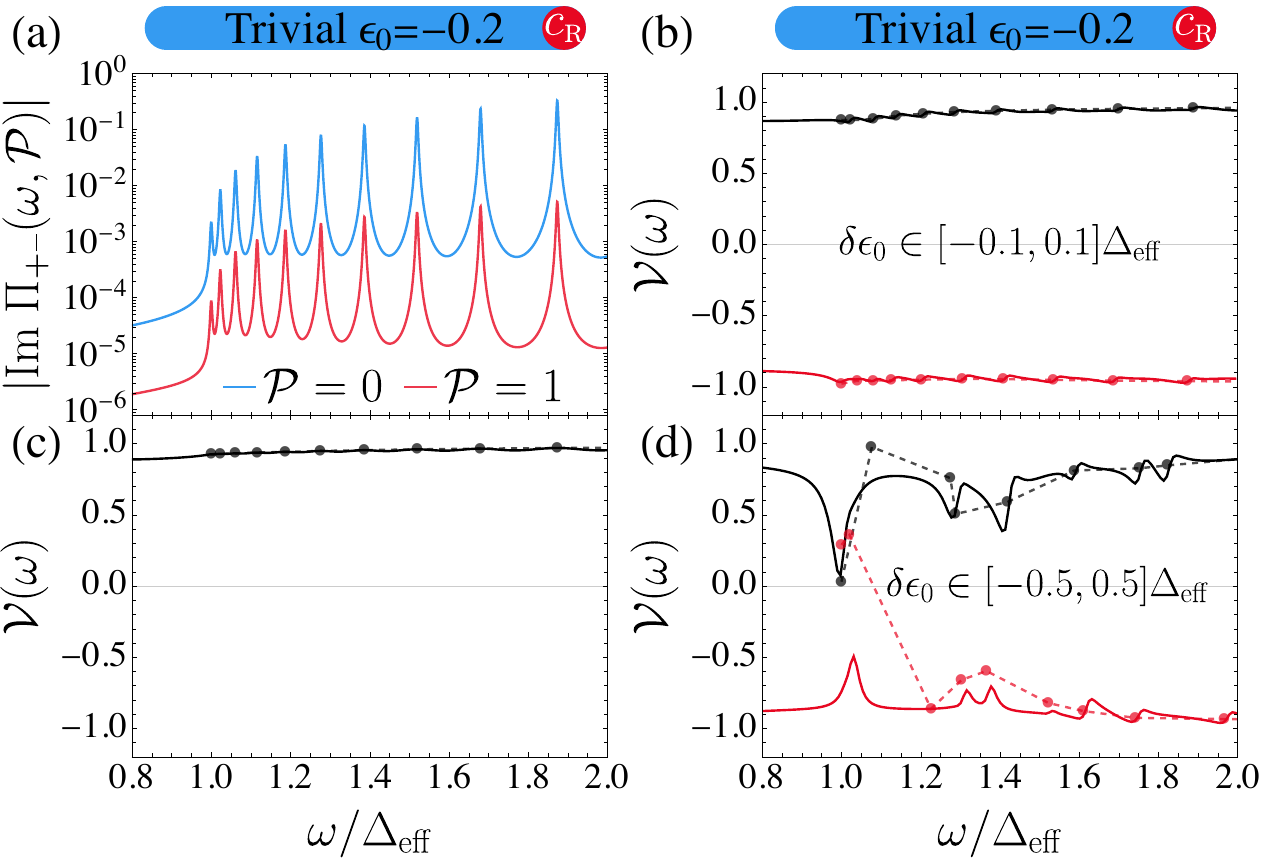}
        \caption{The susceptibility and visibility for another trivial phase $\epsilon_0 = -0.2$. The left panels (a) and (c) are the analogs of Fig.~\ref{Fig:Susceptibility}, while (b) and (d) show the visibility under disorders similar to Fig.~\ref{Fig:Disorders}.}
        \label{Fig:AnotherTrivial}
    \end{minipage}
\end{figure}

To investigate how the total spin expectation value $\langle\tilde{\Sigma}_z\rangle$ differs between the Majorana zero mode and the trivial zero mode in Fig.~\ref{Fig:TotalSpin}, we plot the site-dependent spin expectation values $\langle\tilde{\sigma}_z\rangle$ along the $z$ axis for $N=30$ in Fig.~\ref{Fig:SiteSpin}. We find that $\langle\tilde{\sigma}_z\rangle$ almost remains the same value for two parities at each site in the topological phase. On the other hand, since we fine tune the last site to create the trivial zero mode, $\langle\tilde{\sigma}_z\rangle$ shows a distinct discrepancy between two parities at the end of the chain.

As a supplement to Fig.~\ref{Fig:Disorders}, we demonstrate some specific numerical realizations (without average) of the visibility $\mathcal{V}(\omega)$ in the presence of random onsite disorders $\delta\epsilon_0$ in Fig.~\ref{Fig:DisordersSupp}. For a moderate disorder $\delta\epsilon_0 \in [-0.1, 0.1] \Delta_\mathrm{eff}$ shown in Fig.~\ref{Fig:DisordersSupp}(a)(b), we obtain similar patterns as the ones depicted in Fig.~\ref{Fig:Disorders}(a)(b). When the disorder becomes as large as $\delta\epsilon_0 \in [-0.5, 0.5] \Delta_\mathrm{eff}$, the frequency dependence of the visibility $\mathcal{V}(\omega)$ can be strongly affected in some peculiar realizations. 

We also append the susceptibility and visibility of another trivial phase by setting $k_\textsc{f}a = 5.9 \pi$ and $\epsilon_0 = -0.2$ in Fig.~\ref{Fig:AnotherTrivial}(a)(c), as well as its frequency dependence of the visibility $\mathcal{V}(\omega)$ under random onsite disorders $\delta\epsilon_0$ in Fig.~\ref{Fig:AnotherTrivial}(b)(d). Compared to the $\epsilon_0 = 0.13$ trivial phase shown in Fig.~\ref{Fig:Susceptibility}(b)(d), we found that $\mathcal{P} = 0$ completely dominates the susceptibility over $\mathcal{P} = 1$ in the low-energy regime in Fig.~\ref{Fig:AnotherTrivial}(a). Hence, if the system is in the topologically trivial phase, the visibility in Fig.~\ref{Fig:AnotherTrivial}(c) is close to 1 and can not display a strong oscillation pattern as the one in the topological phase shown in Fig.~\ref{Fig:Susceptibility}(c). When the strength of disorders is moderate, the visibility in Fig.~\ref{Fig:AnotherTrivial}(b) still remains a flat pattern as in Fig.~\ref{Fig:AnotherTrivial}(c). For the realizations with a stronger disorder in Fig.~\ref{Fig:AnotherTrivial}(d), the visibility also shows a similar pattern as in Fig.~\ref{Fig:DisordersSupp}(c). More numerical realizations can be implemented with the source code in Ref.~\cite{Note10}.

\section{Technical details on the overlap integrals}
\label{Appendix:IntegralExpression}

The integral of a function $f(\bd{k})$ in a 2D superconductor can be evaluated by changing the momentum $\bd{k}$ to the angle $\theta$ and the energy dispersion $\varepsilon_\pm$ with respect to the positive and negative helicity:
\begin{equation}
    \int \frac{\mathrm{d} \bd{k}}{(2\pi)^2} f(\bd{k}) = \frac{\nu_\pm}{2\pi} \int \mathrm{d} \varepsilon_\pm \int_0^{2\pi} \mathrm{d} \theta f(\varepsilon_\pm, \theta) \,.
\end{equation}
The normalization factor $\mathcal{N}$ of YSR states in Eq.~\eqref{Eq:ShibaBasis} can be obtained by integrating the un-normalized wavefunction $\ket{\psi} = G(\epsilon_0) \mathcal{H}^{N=1}_\mathrm{el-m} \ket{+} \ket{\uparrow} \equiv G(\epsilon_0) \mathcal{H}^{N=1}_\mathrm{el-m} \ket{\phi}$ in the momentum space:
\begin{align}
    &\mathcal{N}= \bra{\psi}\ket{\psi}
    = \int \frac{\mathrm{d} \bd{k}}{(2\pi)^2} \; \bra{\phi} \mathcal{H}^{N=1}_\mathrm{el-m} G^2(\epsilon_0) \mathcal{H}^{N=1}_\mathrm{el-m} \ket{\phi} \nn\\
    &= \frac{1}{4} \int_0^{2\pi} \mathrm{d} \theta \left[
        \frac{\nu_+}{2\pi} \int \mathrm{d} \varepsilon_+ \bra{\phi} \mathcal{H}^{N=1}_\mathrm{el-m} G_+^2(\epsilon_0) \mathcal{H}^{N=1}_\mathrm{el-m} \ket{\phi} + 
        \frac{\nu_-}{2\pi} \int \mathrm{d} \varepsilon_- \bra{\phi} \mathcal{H}^{N=1}_\mathrm{el-m} G_-^2(\epsilon_0) \mathcal{H}^{N=1}_\mathrm{el-m} \ket{\phi} \right]
    = \frac{(1+\alpha^2)^2}{2 \pi \nu_0 \alpha \Delta} \,.
    \label{Eq:Normalization}
\end{align}
For simplicity, the subscript $\pm$ in the integration variable is omitted in the following context. Similarly, the four integrals used in Eq.~\eqref{Eq:Jmatrix} and Eq.~\eqref{Eq:HeffTwoAngles} are computed as follows \cite{Brydon2015Topological}:
\begin{align}
    I_1^\pm(x)&=\frac{\nu_\pm}{2 \pi} \int \mathrm{d} \varepsilon \int_0^{2 \pi} \mathrm{d} \theta \frac{\varepsilon \mathrm{e}^{\mathrm{i} k_\pm(\varepsilon) x \cos \theta}}{E^2-\varepsilon^2-\Delta^2} \frac{\omega_\textsc{d}^2}{\varepsilon^2+\omega_\textsc{d}^2} \,, \quad 
    I_2^\pm(x)=\frac{\nu_\pm}{2 \pi} \int \mathrm{d} \varepsilon \int_0^{2 \pi} \mathrm{d} \theta \frac{\varepsilon \mathrm{e}^{\mathrm{i} \theta} \mathrm{e}^{\mathrm{i} k_\pm(\varepsilon) x \cos \theta}}{E^2-\varepsilon^2-\Delta^2} \frac{\omega_\textsc{d}^2}{\varepsilon^2+\omega_\textsc{d}^2} \,, \nn \\
    I_3^\pm(x)&=\frac{\nu_\pm}{2 \pi} \int \mathrm{d} \varepsilon \int_0^{2 \pi} \mathrm{d} \theta \frac{\mathrm{e}^{\mathrm{i} k_\pm(\varepsilon) x \cos \theta}}{E^2-\varepsilon^2-\Delta^2} \,, \qquad\qquad\;\;
    I_4^\pm(x)=\frac{\nu_\pm}{2 \pi} \int \mathrm{d} \varepsilon \int_0^{2 \pi} \mathrm{d} \theta \frac{\mathrm{e}^{\mathrm{i} \theta} \mathrm{e}^{\mathrm{i} k_\pm(\varepsilon) x \cos \theta}}{E^2-\varepsilon^2-\Delta^2} \,,
    \label{Eq:IntegralDefinition}
\end{align}
where $k_\pm(\varepsilon)$ is linearized around the Fermi level, namely $k_\pm(\varepsilon) \approx k^\pm_\textsc{f} + \varepsilon/(\hbar v_\textsc{f})$ \cite{Pientka2013Topological}. Different from Ref.~\cite{Brydon2015Topological}, a convergence factor $\omega_\textsc{d}^2/(\varepsilon^2+\omega_\textsc{d}^2)$ is added into $I_{1,2}^\pm(x)$, and $\omega_\textsc{d}$ is the Debye frequency. Using the residue theorem, we obtain \cite{Note10}
\begin{align}
    I_1^\pm(x)
    &= \frac{\pi \nu_\pm \omega_\textsc{d}^2}{E^2-\Delta^2+\omega_\textsc{d}^2}\mathrm{Im}\{J_0[(k^\pm_\textsc{f}+ \mathrm{i} / \xi_\textsc{e})|x|]+ \mathrm{i} H_0[(k^\pm_\textsc{f}+ \mathrm{i} / \xi_\textsc{e})|x|] - J_0[(k^\pm_\textsc{f}+ \mathrm{i} / \xi_\textsc{d})|x|] - \mathrm{i} H_0[(k^\pm_\textsc{f}+ \mathrm{i} / \xi_\textsc{d})|x|]\} \,, \nn \\
    I_2^\pm(x)
    &= \frac{- \mathrm{sgn}(x) \mathrm{i} \pi \nu_\pm \omega_\textsc{d}^2}{E^2-\Delta^2+\omega_\textsc{d}^2} \mathrm{Re}\{ \mathrm{i} J_1[(k^\pm_\textsc{f}+ \mathrm{i} / \xi_\textsc{e})|x|]+H_{-1}[(k^\pm_\textsc{f}+ \mathrm{i} / \xi_\textsc{e})|x|] - \mathrm{i} J_1[(k^\pm_\textsc{f}+ \mathrm{i} / \xi_\textsc{d})|x|] - H_{-1}[(k^\pm_\textsc{f} - \mathrm{i} / \xi_\textsc{d})|x|]\} \,, \nn \\
    I_3^\pm(x)
    &=-\frac{\pi \nu_\pm}{\sqrt{\Delta^2-E^2}} \mathrm{Re}\{J_0[(k^\pm_\textsc{f}+ \mathrm{i} / \xi_\textsc{e} )|x|]+ \mathrm{i} H_0[(k^\pm_\textsc{f}+ \mathrm{i} / \xi_\textsc{e} )|x|]\} \,, \nn \\
    I_4^\pm(x)
    &=-\mathrm{sgn}(x) \frac{ \mathrm{i} \pi \nu_\pm}{\sqrt{\Delta^2-E^2}} \mathrm{Im}\{ \mathrm{i} J_1[(k^\pm_\textsc{f}+ \mathrm{i} / \xi_\textsc{e} )|x|]+H_{-1}[(k^\pm_\textsc{f}+ \mathrm{i} / \xi_\textsc{e} )|x|]\} \,,
    \label{Eq:IntegralExact}
\end{align}
where $J_n$ and $H_n$ are Bessel and Struve functions of order $n$, $\xi_\textsc{e} = \hbar v_\textsc{f}/\sqrt{\Delta^2 - E^2}$ is the superconducting coherence length, and $\xi_\textsc{d} = \hbar v_\textsc{f}/\omega_\textsc{d}$. For large $k^\pm_\textsc{f}$, these integrals have the following asymptotic forms:
\begin{align}
    \label{Eq:IntegralAsymptotic}
    I_1^\pm(x) 
    &\approx \pi \nu_\pm \sqrt{\frac{2 / \pi}{k^\pm_\textsc{f}|x|}} \sin \left(k^\pm_\textsc{f}|x|-\frac{\pi}{4}\right) \mathrm{e}^{-|x| / \xi_\textsc{e}} \,, \qquad \quad
    I_2^\pm(x) \approx \mathrm{sgn}(x) \mathrm{i} \pi \nu_\pm \sqrt{\frac{2 / \pi}{k^\pm_\textsc{f}|x|}} \sin \left(k^\pm_\textsc{f}|x|-\frac{3 \pi}{4}\right) \mathrm{e}^{-|x| / \xi_\textsc{e}} \,, \\
    I_3^\pm(x) 
    &\approx\frac{-\pi \nu_\pm}{\sqrt{\Delta^2-E^2}} \sqrt{\frac{2 / \pi}{k^\pm_\textsc{f}|x|}} \cos \left(k^\pm_\textsc{f}|x|-\frac{\pi}{4}\right) \mathrm{e}^{-|x| / \xi_\textsc{e}} \,, \quad
    I_4^\pm(x) 
    \approx \frac{-\mathrm{sgn}(x)\mathrm{i} \pi \nu_\pm}{\sqrt{\Delta^2-E^2}} \sqrt{\frac{2 / \pi}{k^\pm_\textsc{f}|x|}} \cos \left(k^\pm_\textsc{f}|x|-\frac{3 \pi}{4}\right) \mathrm{e}^{-|x| / \xi_\textsc{e}} \,. \nn
\end{align}
We remark that $I_{1,2}^\pm$ do not have the polynomial terms presented in Ref.~\cite{Brydon2015Topological}, as they are canceled out under the limit $\omega_\textsc{d} \rightarrow \infty$. The Fourier transform of these asymptotic integrals can be obtained by $I(k) = \sum_j I (ja) \mathrm{e}^{\mathrm{i}k j a}$, which enables us to write down $\mathcal{H}_\mathrm{eff}$ [Eq.~\eqref{Eq:HeffTwoAngles}] in the momentum space and thus define the topological invariant $Q$: 
\begin{align}
    \mathcal{H}_\mathrm{eff}(k) &= 
    \begin{bmatrix}
        h_\mathrm{eff}(k) & \Delta_\mathrm{eff}(k) \\
        \Delta_\mathrm{eff}^*(k) & -h_\mathrm{eff}^*(-k)
    \end{bmatrix} + \tau_0\otimes b_\mathrm{eff}(k) \,,\quad
    Q = \mathrm{sgn}[h_\mathrm{eff}(0)h_\mathrm{eff}(\pi/a)] \,.
    \label{Eq:HeffMomentum}
\end{align}
Note that $b_\mathrm{eff}(k) = 0$ if the magnetic impurities are polarized in the $x$-$z$ plane. We refer to the full expressions of the momentum terms in Ref.~\cite{Brydon2015Topological}, as they share the same forms after deleting the polynomial terms.

\section{Technical details of numerical calculations}
\label{Appendix:NumericalDetails}

We use the asymptotic form of integrals in Eq.~\eqref{Eq:IntegralAsymptotic} to calculate the $J_\textsc{e} (\bd{r})$ [Eq.~\eqref{Eq:Jmatrix}] and the spectrum of the effective Hamiltonian $\mathcal{H}_\mathrm{eff}$ [Eq.~\eqref{Eq:HeffPlane}]. The phase diagram in Fig.~\ref{Fig:Scheme}(b) is determined by the topological invariant in Eq.~\eqref{Eq:HeffMomentum}. As we substitute $\Sigma_\nu\rightarrow 2\mathcal{N}\alpha^2(1+\alpha^2)^{-2}\Sigma_\nu = \widetilde{\Sigma}_\nu$ in Eq.~\eqref{el-m}, the total spin shown in Fig.~\ref{Fig:TotalSpin} is indeed calculated by the wavefunctions normalized by $(1+\alpha^2)^{2}/2\alpha^2$ instead of $\mathcal{N}$. Hence we do not need to specify the numerical value of $J$ and $S$. Since we focus on the low-energy modes, we set $E \approx 0$ in the $J_\textsc{e} (\bd{r})$ of Eq.~\eqref{Eq:ShibaWaveFunctionOverSpace}, as well as the coherence length $\xi_\textsc{e} = \xi_0$ in Eq.~\eqref{Eq:IntegralAsymptotic}. We use Eq.~\eqref{Eq:SusceptibilityFull} to calculate the susceptibility (including the bulk-bulk contributions) and its corresponding visibility, where we set $\eta = 2\times 10^{-4}$ and $\mathrm{d}\omega = 0.2\eta$ in Fig.~\ref{Fig:Susceptibility}.

The trivial zero mode is obtained by fine tuning $\epsilon'_0$ at the last site, while $\epsilon_0$ at other sites remains unperturbed in Fig.~\ref{Fig:TotalSpin} and Fig.~\ref{Fig:Susceptibility}. For the trivial phase with $\epsilon_0 = +0.13$, we set $\epsilon'_0 = -0.0668511$ since it is the point where the Pfaffian of $\mathcal{H}_\mathrm{eff}$ changes sign. Similarly, we set $\epsilon'_0 = 0.0334011$ for the trivial phase with $\epsilon_0 = -0.2$ in Fig.~\ref{Fig:AnotherTrivial}. The reason why we choose these two $\epsilon_0$ values for the two trivial phases is that their $\Delta_{\rm eff}$ are comparable to the one in the topological phase, which ensures a fair comparison. As the normalization factor in Eq.~\eqref{Eq:NormalizationFactor} depends on $\alpha = \sqrt{(\Delta-\epsilon_0)/(\Delta+\epsilon_0)}$, we should normalize the YSR wavefunction with the specific value of $\epsilon_0$ at each site. In Fig.~\ref{Fig:Disorders}, we add the disorders randomly site by site in the topological regime. As the zero mode in the trivial phase relies on the fine tuning of the last site, the disorders are randomly added to all sites except the last one. In contrast to the robust Majorana zero mode, the added disorders will slightly enlarge the value of the fine-tuned zero mode. For the trivial zero modes that are lifted to be larger than the resolution, $\eta = 1\times 10^{-3}$ and $\mathrm{d}\omega = 0.5\eta$, we can directly distinguish them from Majorana zero modes by different resonance peaks of the parity-dependent spin susceptibility. To maintain a fair comparison, we focus on the trivial zero modes which are still smaller than the resolution, and differentiate them from Majorana zero modes by the visibility. To calculate the magnonic spectra shown in Fig.~\ref{Fig:MagnonicSpectrum}, we adopt the parameters along the $[1\bar{1}1]$ direction in Fig.~S8 of Ref.~\cite{Kuster2022NonMajorana} , namely $J^\mathrm{ex}_1 = -7.47/2$ meV, $J^\mathrm{ex}_2 = -1.95/2$ meV, $D_1 = -0.07/2$ meV, $D_2 = -0.01/2$ meV, $K_z = 0.19\times 2$ meV, $S=2$, and $H=0$. More numerical details can be found in Ref.~\cite{Note10}.

\end{widetext}

\bibliography{shiba_chains}

\end{document}